\documentclass[preprintnumbers, prd, twocolumn, showpacs, floatfix,
preprintnumbers, letterpaper, superscriptaddress,nofootinbib]{revtex4}
\usepackage{graphicx,epsfig}
\usepackage{amsmath}
\usepackage {amssymb}
\usepackage{subfigure}

\usepackage{relsize}

\newcommand{\be}{\begin{equation}}
\newcommand{\ee}{\end{equation}}
\newcommand{\bea}{\begin{eqnarray}}
\newcommand{\eea}{\end{eqnarray}}
\newcommand{\nn}{\nonumber}

\begin{document}

\title{Teleparallel equivalent of Gauss-Bonnet gravity and its modifications}

\author{Georgios Kofinas}
\email{gkofinas@aegean.gr} \affiliation{Research Group of Geometry,
Dynamical Systems and Cosmology,
Department of Information and Communication Systems Engineering\\
University of the Aegean, Karlovassi 83200, Samos, Greece}

\author{Emmanuel N. Saridakis}
\email{Emmanuel\_Saridakis@baylor.edu}
\affiliation{Physics Division,
National Technical University of Athens, 15780 Zografou Campus,
Athens, Greece} \affiliation{Instituto de F\'{\i}sica, Pontificia
Universidad de Cat\'olica de Valpara\'{\i}so, Casilla 4950,
Valpara\'{\i}so, Chile}


\begin{abstract}
Inspired by the teleparallel formulation of General Relativity, whose
Lagrangian is the torsion invariant $T$, we have constructed the teleparallel
equivalent of Gauss-Bonnet gravity in arbitrary dimensions. Without imposing
the  Weitzenb{\"{o}}ck connection, we have extracted the torsion invariant
$T_G$, equivalent (up to boundary terms) to the Gauss-Bonnet term $G$.
$T_{G}$ is constructed by the vielbein and the connection, it contains
quartic powers of the torsion tensor, it is diffeomorphism and Lorentz
invariant, and in four dimensions it reduces to a topological invariant as
expected. Imposing the Weitzenb{\"{o}}ck connection, $T_G$ depends only on
the vielbein, and this allows us to consider a novel class of modified
gravity theories based on $F(T,T_G)$, which is not spanned by the class of
$F(T)$ theories, nor by the $F(R,G)$ class of curvature modified gravity.
Finally, varying the action we extract the equations of motion for $F(T,T_G)$
gravity.
\end{abstract}


\maketitle

\section{Introduction}
\label{Introduction}

The central foundation of Einstein's gravitational ideas is that gravity is
described through geometry. In his first complete gravitational theory,
General Relativity (GR), he made the additional {\it{assumption}} that geometry
should be described only by curvature, setting torsion to zero, along with
vanishing non-metricity
\cite{Weinberg:2008}. Technically, this is achieved by assuming the
connection to be symmetric in coordinate frame, that is using the Levi-Civita connection. In this
framework one can construct the curvature (Riemann) tensor which carries
all the information of the geometry, and thus of the gravitational field too,
and then, by suitable contractions the simplest (Ricci) scalar $R$ can be constructed,
which contains up to second-order derivatives in the metric.
This Ricci scalar is exactly the Einstein-Hilbert Lagrangian,
whose action gives rise to the Einstein field equations through variation in
terms of the metric.

However, some years later, it was Einstein himself that realized that the
same gravitational equations could arise by a different geometry,
characterized not by curvature but by torsion \cite{ein28}. Technically, this
is achieved by {\it{assuming}} that the antisymmetric piece of the
connection is not vanishing, that is
using the Weitzenb{\"{o}}ck connection. In this framework, one can construct
the torsion tensor, which carries all the information of the geometry and
therefore of the gravitational field, and then simple scalars can be constructed
which contain up to first-order vierbein derivatives.
Finally, one can take a specific combination of these scalars and define
the ``torsion'' scalar $T$, which will be used as the gravitational
Lagrangian, demanding its action to give rise to the Einstein gravitational
field equations through variation in terms of the vierbein. Since these equations coincide
with those of General Relativity, Einstein called this alternative
formulation ``Teleparallel Equivalent of General Relativity''
(TEGR).

On the other hand, the non-renormalizability of General Relativity,
string theory consequences, and the need to describe the universe
acceleration, led a huge amount of research towards the modification
of gravity at the classical level. Using General Relativity as the
starting theory, the simplest modification is to generalize the
action using arbitrary functions of the Ricci scalar, resulting to
the so-called $F(R)$ modified gravity
\cite{DeFelice:2010aj,Nojiri:2010wj}, which has the advantage of
being ghost free.
However, one can
construct more complicated generalizations of the Einstein-Hilbert action by
introducing higher-curvature corrections, such as the Gauss-Bonnet term $G$
\cite{Wheeler:1985nh,Antoniadis:1993jc,Nojiri:2005jg} or functions of it
  \cite{Nojiri:2005jg,DeFelice:2008wz}, Lovelock combinations
\cite{Lovelock:1971yv,Deruelle:1989fj}, Weyl combinations
\cite{Mannheim:1988dj}, or
higher spatial-derivatives as in Ho\v{r}ava-Lifshitz gravity
\cite{Horava:2008ih}.

Hence, a question that arises naturally is the following: can we modify
gravity starting from TEGR instead of General Relativity, that is from its
torsional formulation? For the moment, and inspired by the $F(R)$
modification of General Relativity, only the simplest such torsional
modification exists, namely the $F(T)$ paradigm, in which one extends the
teleparallel Lagrangian $T$ to an arbitrary function $F(T)$
\cite{Ferraro:2006jd,Linder:2010py}. Interestingly enough, although
TEGR coincides with General Relativity at the level of equations of motion,
$F(T)$ does not coincide with $F(R)$, so $F(T)$ is a novel class of gravitational modification with
no (known) equivalent curvature description. This feature led to a detailed
investigation of its cosmological implications
\cite{Ferraro:2006jd,Linder:2010py,Myrzakulov:2010vz,Wu:2010mn} and
black-hole behavior \cite{Wang:2011xf}.

In this work, we are interested in extending the modification of TEGR
inserting higher-order torsion invariants. In particular, inspired by the
Gauss-Bonnet (GB) modification of General Relativity, we first construct
the Teleparallel Equivalent of Gauss-Bonnet term (TEGB) by finding its ``torsion''
equivalent $T_{G}$, which gives the GB field equations.
Then, we use it in order to formulate a modification of TEGR. As a result,
the modification of TEGR plus the TEGB term does not coincide with the
modification of GR plus the GB term, so it is a novel modification of
gravity with no (known) curvature formulation.

The plan of the work is as follows: In section \ref{TEGRsec} we
review the teleparallel formulation of GR in both the coordinate and
the differential form language. In section \ref{gb} we find the
teleparallel equivalent of GB gravity, while in section \ref{EF} we
derive the equations of motion for the general $F(T,T_{G})$ theory.
Finally, a summary of the obtained results is given in section
\ref{Conclusions} of  conclusions.

\section{Construction of Teleparallel Equivalent of General Relativity}
\label{TEGRsec}

In this section we present the construction of Teleparallel Equivalent of
General Relativity. We follow the detailed and conceptually
more enlightening way of construction, starting from an arbitrary connection
with vanishing curvature and not restricted to the  Weitzenb{\"{o}}ck one \cite{Pereira, Maluf,
Maluf:2013gaa}. The benefit of this is that the quantities
defined are both Lorentz and diffeomorphism invariants. In the same spirit we
 continue in the next section with the procedure of constructing
the Teleparallel Equivalent of the Gauss-Bonnet combination. As usual, in
the end we focus on the Weitzenb{\"{o}}ck connection.

In the whole manuscript we use the following notation: Greek indices $\mu, \nu,$...
run over all coordinates of $D$-dimensional space-time
$1,2,...,D$, while Latin indices $a, b, $... run over the tangent
space $1,2,...,D$. Note that we perform the analysis both in the
coordinate and the form languages. Although in the $f(T)$
literature the former is preferred, going to more complicated expressions,
such as the Gauss-Bonnet term, the latter proves much more convenient.

\subsection{Construction of TEGR in coordinate language}
\label{TEGRcoord}

The dynamical variables in torsional formulation of gravity are the
vielbein field $e_a(x^\mu)$, and the connection 1-forms $\omega^a_{\,\,\,
b}(x^\mu)$ which defines the parallel transportation. In terms of
coordinates, they can be expressed in components as
$e_a=e^{\,\,\, \mu}_a\partial_\mu$ and $\omega^a_{\,\,\,b}=\omega^a_{\,\,\,
b\mu}dx^\mu=\omega^a_{\,\,\,bc}e^c$.
The dual vielbein is defined as $e^a=e^a_{\,\,\, \mu}d x^\mu$.
One can express the commutation relations of the vielbein as
\begin{equation}
[e_{a},e_{b}]=C^{c}_{\,\,\,ab}e_{c}\,,
\label{ghw}
\end{equation}
where $C^{c}_{\,\,\,ab}$ are the structure coefficients functions given by
\begin{equation}
C^{c}_{\,\,\,ab}=e_{a}^{\,\,\,\mu}
e_{b}^{\,\,\,\nu}(e^{c}_{\,\,\,\mu,\nu}-e^{c}_{\,\,\,\nu,\mu})
\label{structurefun}\,,
\end{equation}
and comma denotes differentiation.

One can now define the torsion  tensor, expressed in tangent components  as
\begin{equation}
T^{a}_{\,\,\,bc}=
\omega^{a}_{\,\,\,cb}-\omega^{a}_{\,\,\,bc}-C^{a}_{\,\,\,bc}\,,
\end{equation}
and in ``mixed'' ones as
\begin{equation}
T^{a}_{\,\,\,\mu\nu}=
e^{a}_{\,\,\,\nu,\mu}-e^{a}_{\,\,\,\mu,\nu}+\omega^{a}_{\,\,\,b\mu}e^{b}_{\,\,\,\nu}
-\omega^{a}_{\,\,\,b\nu}e^{b}_{\,\,\,\mu}\,.
\label{torsionbastard}
\end{equation}
Similarly, one can define the curvature tensor as
\begin{eqnarray}
&&\!\!\!\!\!\!\!\!\!
R^{a}_{\,\,\,bcd}\!=\omega^{a}_{\,\,\,bd,c}-\omega^{a}_{\,\,\,
bc,d}+\omega^{e}_{\,\,\,bd}\omega^{a}_{\,\,\,ec}
-\omega^{e}_{\,\,\,bc}\omega^{a}_{\,\,\,ed}  -C^{e}_{\,\,\, cd}
\omega^{a}_{\,\,\, be}\nonumber
\\
&&\!\!\!\!\!\!\!\!\! R^{a}_{\,\,\, b\mu\nu}\!=\omega^{a}_{\,\,\,b\nu,\mu}-
\omega^{a}_{\,\,\,b\mu,\nu}
+\omega^{a}_{\,\,\,c\mu}\omega^{c}_{\,\,\,b\nu}-\omega^{a}_{\,\,\,c\nu}
\omega^{c}_{\,\,\,b\mu}\,.
\label{curvaturebastard}
\end{eqnarray}
Thus, as one can see from (\ref{torsionbastard}) and
(\ref{curvaturebastard}), although the torsion tensor depends on both the
vielbein and the connection, that is
$T^{a}_{\,\,\,\mu\nu}(e^{a}_{\,\,\,\mu},\omega^{a}_{\,\,\,b\mu})$,  the
curvature tensor depends only on the connection, namely
$R^{a}_{\,\,\,b\mu\nu}(\omega^{a}_{\,\,\,b\mu})$.

Additionally, there is an independent object which is the metric tensor $g$.
This allows us to make the vielbein orthonormal $g(e_a,e_b)=\eta_{ab}$, where
$\eta_{ab}=\text{diag}(-1,1,...1)$, and we have the relation
\begin{equation}
\label{metrdef}
g_{\mu\nu} =\eta_{ab}\, e^a_{\,\,\,\mu}  \, e^b_{\,\,\,\nu}.
\end{equation}
Indices $a,b,...$ are raised/lowered with the Minkowski metric $\eta_{ab}$.
Finally, throughout the work we impose zero
non-metricity, i.e. $\eta_{ab|c}=0$, which means
$\omega_{abc}=-\omega_{bac}$,
where $|$ denotes covariant differentiation with respect to the connection
$\omega^{a}_{\,\,\,bc}$.

As it is well known, amongst the infinite connection choices there
is only one that gives vanishing torsion, namely the  Christoffel or
Levi-Civita one $\Gamma^{a}_{\,\,\,b}$, with
$\Gamma_{abc}=\frac{1}{2}(C_{cab}-C_{bca}-C_{abc})$, or inversely
$C_{abc}=\Gamma_{acb}-\Gamma_{abc}$. For clarity, we denote the
curvature tensor corresponding to the Levi-Civita connection as
$\bar{R}^{a}_{\,\,\,bcd}$.
The arbitrary connection $\omega_{abc}$ is then related to the
Christoffel connection $\Gamma_{abc}$ through the relation
\begin{equation}
\omega_{abc}=\Gamma_{abc}+\mathcal{K}_{abc}\,,
\label{omega}
\end{equation}
where
\begin{equation}
\mathcal{K}_{abc}=\frac{1}{2}(T_{cab}-T_{bca}-T_{abc}
)=-\mathcal{K}_{bac}
\end{equation}
is the contorsion tensor. Inversely, one can straightforwardly find that
$T_{abc}=\mathcal{K}_{acb}-\mathcal{K}_{abc}$,
while the ``mixed'' contorsion components write as
$\mathcal{K}^{a}_{\,\,\,\mu\nu}=
-\frac{1}{2}(T^{a}_{\,\,\,\mu\nu}+T^{b}_{\,\,\,\mu\lambda}e_{b\nu}
e^{a\lambda}
+T^{b}_{\,\,\,\nu\lambda}e_{b\mu}e^{a\lambda})$,
that is $\mathcal{K}^{a}_{\,\,\,\mu\nu}(e^{a}_{\,\,\,\mu},
\omega^{a}_{\,\,\,b\mu})$.

As long as the vielbein $e^{a}_{\,\,\,\mu}$ and the connection
$\omega^{a}_{\,\,\,b\mu}$ remain independent from each other, the
Einstein-Hilbert Lagrangian density $eR$ (with
$R=e^{a\mu}e^{b\nu}R_{ab\mu\nu}$ the Ricci scalar and
$e=\det{(e^{a}_{\,\,\,\mu})}=\sqrt{|g|}$)
is a function of $e^{a}_{\,\,\,\mu},\omega^{a}_{\,\,\,b\mu}$, and thus
a first-order formulation is needed.

If we now calculate the Ricci scalar $R$ corresponding to the arbitrary
connection, and the Ricci scalar $\bar{R}$ corresponding to the Levi-Civita
connection, they are found to be related through
\begin{eqnarray}
&&eR=e\bar{R}+\frac{1}{4}e\Big(T^{\mu\nu\lambda}T_{\mu\nu\lambda}+2T^{
\mu\nu\lambda}
T_{\lambda\nu\mu}-4T_{\nu}^{\,\,\,\nu\mu}T^{\lambda}_{\,\,\,\lambda\mu}
\Big)\nonumber\\
&&\ \ \ \ \ \ \ \ \ \ \ \  -2(eT_{\nu}^{\,\,\,\nu\mu})_{,\mu}\nonumber\\
&&\ \ \  \,
\ = e\bar{R}+eT -2(eT_{\nu}^{\,\,\,\nu\mu})_{,\mu}\,,
\label{Ricci}
\end{eqnarray}
where we have defined
\begin{eqnarray}
T&=&\frac{1}{4}T^{\mu\nu\lambda}T_{\mu\nu\lambda}+\frac{1}{2}T^{\mu\nu\lambda}
T_{\lambda\nu\mu}-T_{\nu}^{\,\,\,\nu\mu}T^{\lambda}_{\,\,\,\lambda\mu}.
\label{Tquad}
\end{eqnarray}
Since
$e^{-1}(eT_{\nu}^{\,\,\,\nu\mu})_{,\mu}=T_{\nu\,\,\,\,\,\,\,\,;\mu}^{\,\,\,\nu\mu}$,
where $;$ denotes covariant differentiation with respect to the Christoffel connection,
 equation (\ref{Ricci}) is also written as
\begin{equation}
R=\bar{R}+T-2T_{\nu\,\,\,\,\,\,\,\,;\mu}^{\,\,\,\nu\mu}\,.
\label{erg}
\end{equation}
We mention that the quadratic quantity $T$ is diffeomorphism invariant since
$T_{\mu\nu\lambda}$ is a tensor under coordinate transformations.
Additionally, $T$  is local Lorentz invariant, since $T_{abc}$ is a Lorentz
tensor.

One can now introduce the concept of teleparallelism by imposing the condition
of vanishing Lorentz curvature
\begin{eqnarray}
R_{\,\,\,bcd}^{a}=0,
\end{eqnarray}
which holds in all frames. One way to realize this condition is by assuming the
Weitzenb{\"{o}}ck connection $\tilde{\omega}_{\,\,\,\mu\nu}^{\lambda}$ which
is defined in terms of the vielbein $e_{a}^{\,\,\,\mu}$ in
all coordinate frames as
\begin{eqnarray}
\tilde{\omega}_{\,\,\,\mu\nu}^{\lambda}=e_{a}^{\,\,\,\lambda}e^{a}_{\,\,\,\mu
, \nu } .
\label{Weinzdef}
\end{eqnarray}
Due to its inhomogeneous transformation law this connection has tangent-space
components $\tilde{\omega}_{\,\,\,bc}^{a}=0$, and then, the corresponding curvature
components are indeed $\tilde{R}_{\,\,\,bcd}^{a}=0$ (tildes denote the quantities calculated
using  the Weitzenb{\"{o}}ck connection). Note that
$e_{a\,\,\,\,|\nu}^{\,\,\,\mu}=0$,
and thus the vielbein $e_{a}^{\,\,\,\mu}$ is autoparallel with respect to
the connection $\tilde{\omega}_{\,\,\,\mu\nu}^{\lambda}$. The corresponding
torsion tensor is related to the structure coefficients, the contorsion
tensor or the Weitzenb{\"{o}}ck connection, through
\begin{eqnarray}
&&\tilde{T}^{a}_{\,\,\,\mu\nu}=e^{a}_{\,\,\,\nu,\mu}-e^{a}_{\,\,\,\mu,\nu}
=-C^{a}_{\,\,\,bc}e^{b}_{\,\,\,\mu}e^{c}_{\,\,\,\nu}\\
&&\tilde{T}^{a}_{\,\,\,bc}=-C^{a}_{\,\,\,bc}=\tilde{\mathcal{K}}^{a}_{\,\,\,
cb } -\tilde{\mathcal{K}}^{a}_{\,\,\,bc}\\
&&\tilde{T}^{\lambda}_{\,\,\,\mu\nu}=\tilde{\omega}^{\lambda}_{\,\,\,\nu\mu}-
\tilde{\omega}^{\lambda}_{\,\,\,\mu\nu}\,,
\end{eqnarray}
while (\ref{omega}) simplifies to
\begin{eqnarray}
&&\Gamma_{abc}=-\tilde{\mathcal{K}}_{abc}.
\end{eqnarray}

Now inserting the condition $R^{a}_{\,\,\,bcd}=0$ into the general
expression (\ref{Ricci}), we   obtain
\begin{eqnarray}
&&e\bar{R}=-eT+2(eT_{\nu}^{\,\,\,\nu\mu})_{,\mu}\,,
\label{Ric}
\end{eqnarray}
or equivalently
\begin{equation}
\bar{R}=-T+2T_{\nu\,\,\,\,\,\,\,\,;\mu}^{\,\,\,\nu\mu}\,.
\label{erg}
\end{equation}

As we observe the Lagrangian density $e\bar{R}$ of General Relativity
(that is the one calculated with the Levi-Civita connection) differs
from the torsion density $-eT$ only by a total derivative. Therefore, one can
immediately deduce that the General Relativity action
\begin{eqnarray}
S_{EH}=\frac{1}{2\kappa_{D}^{2}}\int_{M}d^{D}\!x\,e\,\bar{R},
\label{GRaction}
\end{eqnarray}
is equivalent (up to boundary terms) to the action
\begin{eqnarray}
S_{Tel}^{(1)}[e^{a}_{\,\,\,\mu},\omega^{a}_{\,\,\,b\mu}]
&\!\!=\!\!&-\frac{1}{2\kappa_{D}^{2}}\int_{M}\!\!d^{D}\!x\,
e\,T\nn\\
&\!\!=\!\!&-\frac{1}{8\kappa_{D}^{2}}\int_{M}\!\!d^{D}\!x\,
e\,\Big(T^{abc}T_{abc}+2T^{abc}
T_{cba}\nonumber\\
&&\ \ \ \ \ \ \ \ \ \ \ \ \ \ \ \ \ \ \ \ \
\ \ -4T_{b}^{\,\,\,ba}T^{c}_{\,\,\,ca}\Big)
\label{tele}
\end{eqnarray}
($\kappa_D^2$ is the $D$-dimensional gravitational constant). Indeed,
varying (\ref{tele}) with respect to the vielbein we get equations which
contain up to $e^{a}_{\,\,\,\mu,\nu\lambda}$, $\omega^{a}_{\,\,\,b\mu,\nu}$,
and imposing the teleparallel condition these equations coincide with the
Einstein field equations as they arise varying
(\ref{GRaction}) with respect to the metric \cite{Maluf:2013gaa}.

If the  Weitzenb{\"{o}}ck connection (\ref{Weinzdef}) is adopted, then the
teleparallel action  (\ref{tele}) becomes a functional only of the vielbein,
which is denoted for clarity as $S_{tel}^{(1)}[e^{a}_{\,\,\,\mu}]$ and has
the same functional form as (\ref{tele}), but with tilde quantities.
Varying $S_{tel}^{(1)}[e^{a}_{\,\,\,\mu}]$ with respect to the vielbein gives
again the Einstein field equations. That is why the constructed theory  in which one
uses torsion to describe the gravitational field, under the teleparallelism
condition, was named by Einstein as Teleparallel Equivalent of General
Relativity \footnote{The normalization of the actions $S_{Tel}^{(1)}$,
$S_{tel}^{(1)}$ has been defined such that
$S_{Tel}^{(1)}=S_{tel}^{(1)}=S_{EH}$.}.
Note that now $\tilde{T}$ still remains diffeomorphism invariant,
while the Lorentz invariance has been lost since we have chosen specific
class of frames. The equations of motion, being the Einstein equations, are still Lorentz covariant.
However, when $T$ in the action is replaced by a general function $f(T)$,
the new equations of motion under Lorentz rotations of the vielbein will not be covariant
(although they are form-invariant).
This is not a deficit (it is a sort of analogue of gauge fixing in gauge theories),
and the theory, although not Lorentz covariant, is meaningful.
Not all vielbeins will be solutions of the new equations, and those which solve the
equations will determine the metric uniquely.


An interesting feature of the above analysis is that in (\ref{erg}) the
Lagrangian $\bar{R}$ has been expressed in terms of torsion through a splitting
into the Lorentz and diffeomorphism invariant term $-T$, containing at most
first
order derivatives in the fields $e^{a}_{\,\,\,\mu},\omega^{a}_{\,\,\,b\mu}$, plus
a total divergence  also Lorentz and diffeomorphism invariant containing the second
order derivatives of $e^{a}_{\,\,\,\mu}$. Note that the Riemann tensor
$\bar{R}^{\mu}_{\,\,\,\nu\rho\sigma}=\Gamma^{\mu}_{\,\,\,\nu\sigma,\rho}
-\Gamma^{ \mu}_{ \,\, \, \nu\rho ,\sigma }
+\Gamma^{\tau}_{\,\,\,\nu\sigma}\Gamma^{\mu}_{\,\,\,\tau\rho}
-\Gamma^{\tau}_{\,\,\,\nu\rho}\Gamma^{\mu}_{\,\,\,\tau\sigma}$
is a sum of the first-order in $e^{a}_{\,\,\,\mu}$ terms
$\Gamma^{\tau}_{\,\,\,\nu\sigma}\Gamma^{\mu}_{\,\,\,\tau\rho}
-\Gamma^{\tau}_{\,\,\,\nu\rho}\Gamma^{\mu}_{\,\,\,\tau\sigma}$,  plus the
second-order total divergence terms
$\Gamma^{\mu}_{\,\,\,\nu\sigma,\rho}-\Gamma^{\mu}_{\,\,\,\nu\rho,\sigma}$. A
similar splitting occurs for the Lagrangian density $e\bar{R}$,
known as the ``gamma-gamma'' form \cite{LandauLif}, however in that case,
the first-order terms as well as the total divergence terms are not
diffeomorphism
invariant. Hence, the teleparallel splitting provides an advantage since the
diffeomorphism invariance is maintained in the separate terms.

\subsection{Construction of TEGR in differential form language}
\label{TEGRform}

Let us now repeat the presentation of the previous subsection in differential form
language. We will need the completely antisymmetric symbol $\epsilon_{a_{1}...a_{D}}$,
which has $\epsilon_{1...D}=1$, while the contravariant components
$\epsilon^{a_{1}...a_{D}}=\eta^{a_{1}b_{1}}...\eta^{a_{D}b_{D}}\epsilon_{b_{1
}...b_{D}}$ have $\epsilon^{1...D}=-1$. The dynamical variables are the vielbein
$e^{a}$ and the connection 1-forms $\omega^{a}_{\,\,\,b}$, with $\omega_{ab}=-\omega_{ba}$
due to the vanishing non-metricity. One can express the commutation relations (\ref{ghw})
in terms of the dual vielbein as
\begin{equation}
de^{a}=-\frac{1}{2}C^{a}_{\,\,\,bc}e^{b}\wedge e^{c},
\end{equation}
where $\wedge$ denotes the wedge product.

One can now define the torsion 2-form as
\begin{equation}
T^{a}=de^{a}+\omega^{a}_{\,\,\,b}\wedge e^{b}=\frac{1}{2}
T^{a}_{\,\,\,bc}e^{b}\wedge e^{c},
\label{jil}
\end{equation}
and the curvature 2-form as
\begin{equation}
\mathcal{R}^{a}_{\,\,\,b}=
d\omega^{a}_{\,\,\,b}+\omega^{a}_{\,\,\,c}\wedge\omega^{c}_{\,\,\,b}=\frac{1}
{2}R^{a}_{\,\,\,bcd}e^{c}\wedge e^{d}.
\end{equation}
The curvature 2-form corresponding to  $\Gamma^{a}_{\,\,\,b}$ is denoted  by
$\mathcal{\bar{R}}^{a}_{\,\,\,b}$. The arbitrary connection
$\omega^{a}_{\,\,\,b}$ is then related to $\Gamma^{a}_{\,\,\,b}$ through the relation
\begin{equation}
\mathcal{K}_{ab}
=-\mathcal{K}_{ba}=\omega_{ab}-\Gamma_{ab}=\mathcal{K}_{abc}e^c,
\end{equation}
where $\mathcal{K}_{ab}$ is the contorsion 1-form. Inversely, one can straightforwardly
find that $T^{a}=\mathcal{K}^{a}_{\,\,\,b}\wedge e^{b}$. Finally, note that under the
Weitzenb{\"{o}}ck connection the previous relation simplifies to
\begin{eqnarray}
&&\Gamma_{ab}=-\tilde{\mathcal{K}}_{ab}.
\end{eqnarray}

The  action of  General Relativity  is written in terms of the connection
$\Gamma^{a}_{\,\,\,b}$ as
\begin{eqnarray}
S_{EH}=\frac{1}{2\kappa_{D}^{2}}\int_{M}\mathcal{\bar{L}}_{1},
\end{eqnarray}
where
\begin{equation}
\mathcal{\bar{L}}_{1}= \frac{1}{(D-2)!}
\epsilon_{a_{1}...a_{D}}\mathcal{\bar{R}}^{a_{1}a_{2}}\wedge
e^{a_{3}}\wedge...\wedge e^{a_{D}}=\bar{R}\ast\!1,
\end{equation}
with $\ast $ denoting the Hodge dual operator. If we now calculate the Lagrangian
$\mathcal{L}_{1}$ corresponding to the arbitrary connection $\omega^{a}_{\,\,\,b}$,
it is related to $\mathcal{\bar{L}}_{1}$ through
\begin{eqnarray}
&&(D-2)!\,\mathcal{L}_{1}=(D-2)!\,\mathcal{\bar{L}}_{1}
\nonumber\\
&&\ \ \ \ \ \ \ \ \ \ \ \ \ \ \ \
\ \ \ +d(\epsilon_{a_{1}...a_{D}}\mathcal{K}^{a_{1}a_{2}}\wedge
e^{a_{3}}\wedge...\wedge e^{a_{D}})\nonumber\\
&&\ \ \ \ \ \ \ \ \ \ \ \ \ \ \ \
\ \ \
+\epsilon_{a_{1}...a_{D}}\mathcal{K}^{a_{1}a_{2}}\wedge
d(e^{a_{3}}\wedge...\wedge e^{a_{D}})\nn\\
&&\ \ \ \ \ \ \ \ \ \ \ \ \ \ \ \
\ \ \ +\epsilon_{a_{1}...a_{D}}
\left(\Gamma^{a_{1}}_{\,\,\,\,c}\wedge\mathcal{K}^{ca_{2}}
+\mathcal{K}^{a_{1}}_{\,\,\,\,c}\wedge\Gamma^{ca_{2}}\right. \nonumber\\
&&\left. \ \ \ \ \ \ \ \ \ \ \ \ \ \ \ \
\ \ \ \ \  +
\mathcal{K}^{a_{1}}_{\,\,\,\,c}\wedge\mathcal{K}^{ca_{2}}\right)\wedge
e^{a_{3}}\wedge...\wedge e^{a_{D}},
\label{formRicci1}
\end{eqnarray}
which after some cancelations provides the analogue of (\ref{Ricci})
\begin{eqnarray}
&&\!\!\!\!\!\!\mathcal{L}_{1}\!=\!\mathcal{\bar{L}}_{1}+\frac{1}{(D-2)!}\epsilon_{a_{1}...a_{D}}
\mathcal{K}^{a_{1}}_{\,\,\,\,c}\wedge\mathcal{K}^{ca_{2}}\wedge
e^{a_{3}}\wedge...\wedge e^{a_{D}}
\nonumber\\
&&\ \ \ \ \ \ \ +\frac{1}{(D-2)!}
d(\epsilon_{a_{1}...a_{D}}\mathcal{K}^{a_{1}a_{2}}\wedge e^{a_{3}}\wedge...\wedge
e^{a_{D}}).
\label{formRicci}
\end{eqnarray}

Finally, imposing the teleparallel condition $\mathcal{R}^{ab}=0$, we get the
analogue of (\ref{Ric})
\begin{eqnarray}
&& \!\!\!\!\!\!\!\!\!\!\!\!\!\!\mathcal{\bar{L}}_{1}\!=\!-\mathcal{T}\!-\!\frac{1}{(D\!-\!2)!}
d(\epsilon_{a_{1}...a_{D}}\mathcal{K}^{a_{1}a_{2}}\wedge e^{a_{3}}\wedge...\wedge
e^{a_{D}}),
\label{formRic}
\end{eqnarray}
where
\begin{eqnarray}
\mathcal{T}&\!\!=\!\!&\frac{1}{(D-2)!}\epsilon_{a_{1}...a_{D}}
\mathcal{K}^{a_{1}}_{\,\,\,\,c}\wedge\mathcal{K}^{ca_{2}}\wedge
e^{a_{3}}\wedge...\wedge e^{a_{D}}\nn\\
&\!\!=\!\!&T\,\,e^{1}\wedge ... \wedge e^{D}
\label{tosrionscalarform}
\end{eqnarray}
is the TEGR volume form, and
\begin{equation}
T=\mathcal{K}^{abc}\mathcal{K}_{cba}
-\mathcal{K}^{ca}_{\,\,\,\,\,a}\mathcal{K}_{cb}^{\,\,\,\,\,b}
\label{huy}
\end{equation}
the corresponding scalar. Ignoring the boundary term in (\ref{formRic})
we obtain again the teleparallel action (\ref{tele}) of Einstein gravity
\begin{equation}
S_{Tel}^{(1)}=-\frac{1}{2\kappa_{D}^{2}}\int_{M}\mathcal{T}
=-\frac{1}{2\kappa_{D}^{2}}\int_{M}\!\!d^{D}\!x\,
e\,T\,.
\label{formtele}
\end{equation}

In order to obtain the above results it is much more powerful to
introduce the covariant exterior differential $D$ of the connection $\omega_{ab}$
acting on a set of $p-$forms $\Phi^{a}_{b}$ as
$D\Phi^{a}_{b}=d\Phi^{a}_{b}+\omega^{a}_{\,\,\,c}\wedge\Phi^{c}_{b}-(-1)^{p}
\Phi^{a}_{c}\wedge\omega^{c}_{\,\,\,b}$. Similarly, the differential
$\bar{D}$ is defined for the connection $\Gamma_{ab}$. Then,
$\mathcal{R}^{ab}=\mathcal{\bar{R}}^{ab}+\bar{D}\mathcal{K}^{ab}+\mathcal{K}^
{a}_{\,\,\,c}\wedge\mathcal{K}^{cb}$,
$T^{a}=De^{a}$, $DT^{a}=\mathcal{R}^{a}_{\,\,\,b}\wedge e^{b}$,
$D\mathcal{R}^{a}_{\,\,\,b}=0$,
$D^{2}\Phi^{a}_{b}=\mathcal{R}^{a}_{\,\,\,c}\wedge\Phi^{c}_{b}-\Phi^{a}_{c}
\wedge\mathcal{R}^{c}_{\,\,\,b}$.
Since it is $\bar{D}e^{a}=0$, we get immediately equation (\ref{formRicci}).
This is the method that will be followed in the next section.

\section{Construction of Teleparallel Equivalent of Gauss-Bonnet term}
\label{gb}

In this section we will construct the Teleparallel Equivalent of the
Gauss-Bonnet gravity. We will follow the procedure of the construction of
TEGR described above, based on the corresponding action. The central
strategy of the previous section was to express the curvature scalar $R$
corresponding to a general connection  as the curvature scalar $\bar{R}$
corresponding to Levi-Civita connection, plus terms arising from the torsion
tensor. Then, by imposing the teleparallelism condition $R^{a}_{\,\,\,bcd}=0$,
we acquire  that $\bar{R}$ is equal to a torsion scalar plus a total derivative,
namely relation (\ref{Ric}). This torsion scalar provides  the Teleparallel
Equivalent of General Relativity, in a sense that if one uses it as a Lagrangian,
the same exactly equations with General Relativity are obtained.

In this section we follow the same steps to re-express the Gauss-Bonnet
combination
\begin{equation}
G=R^{2}-4R_{\mu\nu}R^{\mu\nu}+R_{\mu\nu\kappa\lambda}R^{\mu\nu\kappa\lambda}.
\label{GBdef}
\end{equation}
However, for convenience we will use the form language
which leads to simple expressions  compared to the coordinate description.
The action of Gauss-Bonnet gravity in terms of
the Levi-Civita connection  is
\begin{equation}
S_{GB}=\frac{1}{2\kappa_{D}^{2}}\int_{M}\mathcal{\bar{L}}_{2},
\end{equation}
where
\begin{eqnarray}
&&
\mathcal{\bar{L}}_{2}=\frac{1}{(D-4)!}\epsilon_{a_{1}...a_{D}}\mathcal{\bar{
R}}^{a_{1}a_{2}}\wedge
\mathcal{\bar{R}}^{a_{3}a_{4}}\wedge e^{a_{5}}\wedge...\wedge
e^{a_{D}}\nonumber\\
&&\ \  \ \
=\bar{G}\ast\!1.
\end{eqnarray}

The corresponding Lagrangian when $\mathcal{\bar{R}}^{ab}$ is replaced by
$\mathcal{R}^{ab}$, that is the one that corresponds to an arbitrary
connection $\omega^{a}_{\,\,\,b}$, is denoted by
$\mathcal{L}_{2}$. The relation between $\mathcal{L}_{2}$ and
$\mathcal{\bar{L}}_{2}$ is found to be
\begin{equation}
(D-4)!\,\mathcal{L}_{2}=(D-4)!\,\mathcal{\bar{L}}_{2}+I_{1}+2I_{2}+2I_{3}+2I_
{4}+I_{5}\,,
\label{GBformRicci1}
\end{equation}
where
\begin{eqnarray}
I_{1}&=&\epsilon_{a_{1}...a_{D}}\mathcal{K}^{a_{1}}_{\,\,\,\,\,c}
\wedge\mathcal{K}^{ca_{2}}\wedge
\mathcal{K}^{a_{3}}_{\,\,\,\,\,d}\wedge\mathcal{K}^{da_{4}}\wedge
e^{a_{5}}\wedge...\wedge e^{a_{D}}\nn\\
I_{2}&=&\epsilon_{a_{1}...a_{D}}\mathcal{\bar{R}}^{a_{1}a_{2}}\wedge
\mathcal{K}^{a_{3}}_{\,\,\,\,\,c}\wedge\mathcal{K}^{ca_{4}}\wedge
e^{a_{5}}\wedge...\wedge e^{a_{D}}\nn\\
I_{3}&=&\epsilon_{a_{1}...a_{D}}\bar{D}\mathcal{K}^{a_{1}a_{2}}\wedge
\mathcal{K}^{a_{3}}_{\,\,\,\,\,c}\wedge\mathcal{K}^{ca_{4}}\wedge
e^{a_{5}}\wedge...\wedge e^{a_{D}}\nn\\
I_{4}&=&\epsilon_{a_{1}...a_{D}}\bar{D}\mathcal{K}^{a_{1}a_{2}}\wedge
\mathcal{\bar{R}}^{a_{3}a_{4}}\wedge e^{a_{5}}\wedge...\wedge e^{a_{D}}\nn\\
I_{5}&=&\epsilon_{a_{1}...a_{D}}\bar{D}\mathcal{K}^{a_{1}a_{2}}\wedge
\bar{D}\mathcal{K}^{a_{3}a_{4}}\wedge e^{a_{5}}\wedge...\wedge e^{a_{D}}\,.
\label{I}
\end{eqnarray}
$I_{1}$ is an algebraic term quartic in torsion. Since
$\bar{D}\bar{\mathcal{R}}^{ab}=0$
and $\bar{D}e^{a}=0$, $I_{4}$ is an exact form
\begin{eqnarray}
I_{4}=d(\epsilon_{a_{1}...a_{D}}\mathcal{K}^{a_{1}a_{2}}
\wedge
\mathcal{\bar{R}}^{a_{3}a_{4}}\wedge e^{a_{5}}\wedge...\wedge
e^{a_{D}}).
\end{eqnarray}
Similarly, since $\bar{D}^{2}\mathcal{K}^{ab}=
\mathcal{\bar{R}}^{a}_{\,\,\,c}\wedge\mathcal{K}^{cb}+\mathcal{\bar{R}}^{b}_{
\,\,\,c}\wedge\mathcal{K}^{ac}$, we have
\begin{eqnarray}
&&\!\!\!\!\!\!\!\!\!\!\!\!\!I_{5}=2\epsilon_{a_{1}...a_{D}}\mathcal{K}^{a_{1}
a_{2}} \wedge
\mathcal{\bar{R}}^{a_{3}}_{\,\,\,\,\,c}\wedge\mathcal{K}^{ca_{4}}\wedge
e^{a_{5}}\wedge...\wedge e^{a_{D}}\nn\\
&&
+d(\epsilon_{a_{1}...a_{D}}\mathcal{K}^{a_{1}a_{2}}\wedge
\bar{D}\mathcal{K}^{a_{3}a_{4}}\wedge e^{a_{5}}\wedge...\wedge e^{a_{D}}).
\end{eqnarray}
Therefore,
\begin{eqnarray}
(D-4)!\,\mathcal{L}_{2}=(D-4)!\,\mathcal{\bar{L}}_{2}+I_{1}+2I_{3}+2I_{6}
+dB\,,
\label{GBformRicci2}
\end{eqnarray}
where
\begin{eqnarray}
I_{6}&=&\epsilon_{a_{1}...a_{D}}\mathcal{\bar{R}}^{a_{1}a_{2}}\wedge
\mathcal{K}^{a_{3}}_{\,\,\,\,\,c}\wedge\mathcal{K}^{ca_{4}}\wedge
e^{a_{5}}\wedge...\wedge e^{a_{D}}\nn\\
&&
+\epsilon_{a_{1}...a_{D}}\mathcal{K}^{a_{1}a_{2}}\wedge
\mathcal{\bar{R}}^{a_{3}}_{\,\,\,\,\,c}\wedge\mathcal{K}^{ca_{4}}\wedge
e^{a_{5}}\wedge...\wedge e^{a_{D}}
\label{I2prime}\nn\\
B&=&2\epsilon_{a_{1}...a_{D}}\mathcal{K}^{a_{1}a_{2}}\wedge
\mathcal{\bar{R}}^{a_{3}a_{4}}\wedge e^{a_{5}}\wedge...\wedge e^{a_{D}}\nn\\
&&
+\epsilon_{a_{1}...a_{D}}\mathcal{K}^{a_{1}a_{2}}\wedge
\bar{D}\mathcal{K}^{a_{3}a_{4}}\wedge e^{a_{5}}\wedge...\wedge e^{a_{D}}\,.
\label{B}
\end{eqnarray}
Taking into account that
$\mathcal{\bar{R}}^{ab}+\bar{D}\mathcal{K}^{ab}=\mathcal{R}^{ab}-\mathcal{K}^
{a}_{\,\,\,c}\wedge\mathcal{K}^{cb}$, equation (\ref{GBformRicci2}) is written as
\begin{equation}
(D-4)!\,\mathcal{L}_{2}=(D-4)!\,\mathcal{\bar{L}}_{2}+2J_{0}-I_{1}+2J_{1}+dB
\,,
\label{GBformRicci3af}
\end{equation}
where
\begin{eqnarray}
\!\!\!\!J_{0}&\!\!=\!\!&\epsilon_{a_{1}...a_{D}}\mathcal{R}^{a_{1}a_{2}}
\wedge\mathcal{K}^{a_{3}}_{\,\,\,\,\,c}\wedge
\mathcal{K}^{ca_{4}}\wedge e^{a_{5}}\wedge...\wedge e^{a_{D}}\nn
\label{J0a}\\
\!\!\!\!
J_{1}&\!\!=\!\!&\epsilon_{a_{1}...a_{D}}\mathcal{K}^{a_{1}a_{2}}\wedge\mathcal{\bar{R}}^{a_{3}}_{\,\,\,\,\,c}
\wedge\mathcal{K}^{ca_{4}}\wedge e^{a_{5}}\wedge...\wedge e^{a_{D}}\nn\,.
\label{J1er}
\end{eqnarray}
To finish, from the identity $\bar{D}\mathcal{K}^{a}_{\,\,\,b}=D\mathcal{K}^{a}_{\,\,\,b}
-2\mathcal{K}^{a}_{\,\,\,c}\wedge\mathcal{K}^{c}_{\,\,\,b}$ we get $\bar{\mathcal{R}}^{a}_{\,\,\,b}=
\mathcal{R}^{a}_{\,\,\,b}+\mathcal{K}^{a}_{\,\,\,c}\wedge\mathcal{K}^{c}_{\,\,\,b}-D\mathcal{K}^{a}_{\,\,\,b}$
and substituting into $J_{1}$ we obtain
\begin{equation}
(D-4)!\,\mathcal{L}_{2}\!=\!(D-4)!\,\mathcal{\bar{L}}_{2}+2(J_{0}+\hat{J}_{0})-I_{1}+2J_{2}-2J_{3}+dB
\,,
\label{GBformRicci3jib}
\end{equation}
where
\begin{eqnarray}
\!\!J_{2}&\!\!=\!\!&\epsilon_{a_{1}...a_{D}}\mathcal{K}^{a_{1}a_{2}}\wedge
\mathcal{K}^{a_{3}}_{\,\,\,\,\,c}\wedge\mathcal{K}^{c}_{\,\,\,d}
\wedge\mathcal{K}^{da_{4}}\wedge e^{a_{5}}\wedge...\wedge e^{a_{D}}\nn\\
\!\!J_{3}&\!\!\!=\!\!\!&\epsilon_{a_{1}...a_{D}}
\mathcal{K}^{a_{1}a_{2}}\wedge
D\mathcal{K}^{a_{3}}_{\,\,\,\,\,c}\wedge \mathcal{K}^{ca_{4}}
\!\wedge\!
e^{a_{5}}\!\wedge...\wedge\! e^{a_{D}}\nn\\
&\!\!=\!\!&
\epsilon_{a_{1}...a_{D}}\!\left(
\mathcal{K}^{a_{1}a_{2}}\wedge
d\mathcal{K}^{a_{3}}_{\,\,\,\,\,c}\wedge \mathcal{K}^{ca_{4}}\right.\nn\\
&& \ \ \ \ \ \ \ \ \ \
+\mathcal{K}^{a_{1}a_{2}}\!\wedge\!\omega^{a_{3}}_{\,\,\,\,\,\,c}
\!\wedge\!\mathcal{K}^{c}_{\,\,\,d}\!\wedge\!\mathcal{K}^{da_{4}}\nn\\
&&\left. \ \ \ \ \ \ \ \ \ \
+\mathcal{K}^{a_{1}a_{2}}\!\wedge\!\mathcal{K}^{a_{3}}_{\,\,\,\,\,\,c}
\!\wedge\!\omega^{c}_{\,\,\,d}\!\wedge\!\mathcal{K}^{da_{4}}\right)\!\wedge\!
e^{a_{5}}\!\wedge...\wedge\! e^{a_{D}}\nn\\
\!\!\hat{J}_{0}&\!\!=\!\!&\epsilon_{a_{1}...a_{D}}\mathcal{K}^{a_{1}a_{2}}
\wedge\mathcal{R}^{a_{3}}_{\,\,\,\,\,c}\wedge
\mathcal{K}^{ca_{4}}\wedge e^{a_{5}}\wedge...\wedge e^{a_{D}}\,.\nn
\end{eqnarray}

In order to extract the teleparallel equivalent of GB gravity we set
$\mathcal{R}^{ab}=0$ in (\ref{GBformRicci3jib}) obtaining
\begin{eqnarray}
\mathcal{\bar{L}}_{2}=\mathcal{T}_{G}-\frac{1}{(D-4)!} dB\,,
\label{GBformRiccinm}
\end{eqnarray}
where
\begin{eqnarray}
\mathcal{T}_{G}
\!&\!\!\!=\!\!\!&\frac{1}{(D\!-\!4)!}\epsilon_{a_{1}...a_{D}}\big(\mathcal{K}^{a_{1}}_{\,\,\,\,\,c}
\!\wedge\!\mathcal{K}^{ca_{2}}\!\wedge\!\mathcal{K}^{a_{3}}_{\,\,\,\,\,d}\!\wedge\!
\mathcal{K}^{da_{4}}\nn\\
&&\ \ \ \ \ \ \ \ \ \ \ \ \ \ \ \ \ \
-2\mathcal{K}^{a_{1}a_{2}}\!\wedge\!\mathcal{K}^{a_{3}}_{\,\,\,\,\,c}\!\wedge\!
\mathcal{K}^{c}_{\,\,\,d}\!\wedge\!\mathcal{K}^{da_{4}}\nn\\
&&\ \ \ \ \ \ \ \ \ \ \ \ \ \ \ \ \ \
+2\mathcal{K}^{a_{1}a_{2}}\!\wedge\!
D\mathcal{K}^{a_{3}}_{\,\,\,\,\,c}\!\wedge\!
\mathcal{K}^{ca_{4}}\!\big)\!\!\wedge\!e^{a_{5}}\!\wedge...\wedge\!e^{a_{D}}\nn\\
&\!\!=\!\!&T_{G}\,\,e^{1}\wedge ... \wedge e^{D} \label{pokl}
\end{eqnarray}
is {\textit{the TEGB volume form}}. The corresponding {\textit{scalar}} is
\begin{eqnarray}
&&\!\!\!\!\!\!\!\!\!\!T_{G}=\big(\mathcal{K}^{a_{1}}_{\,\,\,\,ea}
\mathcal{K}^{ea_{2}}_{\,\,\,\,\,\,\,b}\mathcal{K}^{a_{3}}_{\,\,\,\,fc}
\mathcal{K}^{fa_{4}}_{\,\,\,\,\,\,\,d}
-2\mathcal{K}^{a_{1}\!a_{2}}_{\,\,\,\,\,\,\,\,\,\,a}\mathcal{K}^{a_{3}}_{
\,\,\,\,\,eb}
\mathcal{K}^{e}_{\,\,fc}\mathcal{K}^{fa_{4}}_{\,\,\,\,\,\,\,\,d}
\nn\\
&& \ \ \ \ \ \
+2\mathcal{K}^{a_{1}\!a_{2}}_{\,\,\,\,\,\,\,\,\,\,a}\mathcal{K}^{a_{3}}_{
\,\,\,\,\,eb}
\mathcal{K}^{ea_{4}}_{\,\,\,\,\,\,\,\,f}\mathcal{K}^{f}_{\,\,\,cd}\nn\\
&&\ \ \ \ \ \
+2\mathcal{K}^{a_{1}\!a_{2}}_{\,\,\,\,\,\,\,\,\,\,a}\mathcal{K}^{a_{3}}_{
\,\,\,\,\,eb}\mathcal{K}^{ea_{4}}_{\,\,\,\,\,\,\,c|d}\big)\delta^{\,a\,b\,c\,d}_{a_{1}a_{2}a_{3}a_{4}}\,.
\label{poklbn}
\end{eqnarray}
Here, $D\mathcal{K}^{a}_{\,\,\,b}=d\mathcal{K}^{a}_{\,\,\,b}+
\omega^{a}_{\,\,\,c}\wedge\mathcal{K}^{c}_{\,\,\,b}+\mathcal{K}^{a}_{\,\,\,c}\wedge\omega^{c}_{\,\,\,b}=
(\mathcal{K}^{a}_{\,\,\,bc|d}+\frac{1}{2}\mathcal{K}^{a}_{\,\,\,be}T^{e}_{\,\,\,dc})e^{d}\wedge e^{c}$,
the covariant derivative of $\mathcal{K}^{a}_{\,\,\,bc}$ with respect to $\omega^{a}_{\,\,\,bc}$ is
$\mathcal{K}^{a}_{\,\,\,bc|d}=\mathcal{K}^{a}_{\,\,\,bc,d}+\omega^{a}_{\,\,\,ed}\mathcal{K}^{e}_{\,\,\,bc}
-\omega^{e}_{\,\,\,bd}\mathcal{K}^{a}_{\,\,\,ec}-\omega^{e}_{\,\,\,cd}\mathcal{K}^{a}_{\,\,\,be}$,
the $C^{a}_{\,\,\,bc}$ is given by equation (\ref{structurefun}),
and the generalized $\delta$  is the determinant of the Kronecker deltas.

The analogue of equation (\ref{Ric}) is now
\begin{eqnarray}
&&\!\!\!\!\!\!\!\!\!\!\!\!\!\!\!e(\bar{R}^{2}\!-\!4\bar{R}_{\mu\nu}\bar{R}^{\mu\nu}
\!+\!\bar{ R}_{\mu\nu\kappa\lambda}\bar{R}^{\mu\nu\kappa\lambda})
\!=\!eT_{G}\!+\!\text{total diverg.}
\label{TG}
\end{eqnarray}
Obviously, $\mathcal{T}_{G}$ is a Lorentz invariant made out of $e^{a}$, $\omega^{a}_{\,\,\,b}$.
Since in $D=4$ dimensions the GB term $\mathcal{\bar{L}}_{2}^{(D=4)}$ is a
topological invariant, so must be $\mathcal{T}_{G}^{(D=4)}$. Indeed, it is
\begin{equation}
\mathcal{T}_{G}^{(D=4)}=d\big(32\pi^{2}\,\Pi_{2}+B\big)\,,
\label{dftd}
\end{equation}
where
\begin{equation}
\Pi_{2}=-\frac{1}{8\pi^{2}}\epsilon_{abcd}n^{a}(\varepsilon\bar{\mathcal{R}}^{bc}\wedge \bar{D}n^{d}+
\frac{2}{3}\bar{D}n^{b}\wedge \bar{D}n^{c}\wedge \bar{D}n^{d})
\label{chern1}
\end{equation}
is the second Chern form, $n^{a}$ is a unit vector with $n^{a}n_{a}=\varepsilon=\pm 1$,
and $\mathcal{\bar{L}}_{2}^{(D=4)}=32\pi^{2}d\Pi_{2}$.
{\it{Therefore, we have constructed a new Lorentz invariant $\mathcal{T}_{G}$ out
of $e^{a}$, $\omega^{a}_{\,\,\,b}$, containing quartic powers of the torsion tensor, which
in 4 dimensions   becomes a topological invariant.}}


Ignoring the boundary term $B$ in (\ref{GBformRiccinm}) we obtain the
teleparallel action of Gauss-Bonnet gravity
\begin{equation}
S_{Tel}^{(2)}[e^{a},\omega^{a}_{\,\,\,b}]=\frac{1}{2\kappa_{D}^{2}}\int_{M}\mathcal{T}_{G}
=\frac{1}{2\kappa_{D}^{2}}\int_{M}\!\!d^{D}\!x\,
e\,T_{G}\,.
\label{poklc}
\end{equation}
The action $S_{Tel}^{(2)}[e^{a}_{\,\,\,\mu},\omega^{a}_{\,\,\,b\mu}]$
is diffeomorphism and  Lorentz invariant. Beyond
$e^{a}_{\,\,\,\mu}, \,e^{a}_{\,\,\,\mu,\nu}, \,\omega^{a}_{\,\,\,b\mu}$,
which exist in $S_{Tel}^{(1)}$ in (\ref{tele}) too, in
$S_{Tel}^{(2)}$ there appear also
$e^{a}_{\,\,\,\mu,\nu\lambda},\,\omega^{a}_{\,\,\,b\mu,\nu}$, but in a form such that
the equations of motion do not contain higher than second derivatives
in $e^{a}_{\,\,\,\mu}$, as expected from the Gauss-Bonnet term.

Now, choosing the  Weitzenb{\"{o}}ck connection $\omega^{a}_{\,\,\,bc}=0$,
the action (\ref{poklc}) becomes
\begin{eqnarray}
S_{tel}^{(2)}\!\!\!&=&\!\!\!\frac{1}{2(D\!-\!4)!\,\kappa_
{D }^ {2 }}
\!\int_{M}\!\!\!\epsilon_{a_{1}...a_{D}}
\big(\mathcal{K}^{a_{1}}_{\,\,\,\,\,c}\!\wedge\!\mathcal{K}^{ca_{2}}
\!\wedge\!
\mathcal{K}^{a_{3}}_{\,\,\,\,\,d}\!\wedge\!\mathcal{K}^{da_{4}}\nn\\
&& \ \ \ \ \ \ \ \ \ \ \ \ \ \ \
-2\mathcal{K}^{a_{1}a_{2}}\!\wedge\!\mathcal{K}^{a_{3}}_{\,\,\,\,\,c}
\!\wedge\!\mathcal{K}^{c}_{\,\,\,d}
\!\wedge\!\mathcal{K}^{da_{4}}
\nn\\
&& \ \ \ \ \ \ \ \ \ \ \ \ \ \ \
-2\mathcal{K}^{a_{1}a_{2}}\!\wedge\!
\mathcal{K}^{a_{3}}_{\,\,\,\,\,c}\!\wedge\!
d\mathcal{K}^{ca_{4}}\big)\!\!\wedge\! e^{a_{5}}\!\wedge\!...\!\wedge\!
e^{a_{D}}\!.\nn\\
\label{teleGBform}
\end{eqnarray}
Note that the tildes denoting the quantities corresponding to the
Weitzenb{\"{o}}ck connection are omitted for simplicity.
In coordinate language it is
\begin{eqnarray}
\!\!\!\!\!\!S_{tel}^{(2)}\!\!\!&=&\!\!\!
\frac{1}{2\kappa_{D}^{2}}\!\int_{M}\!\!\!d^{D}\!x\,
e\,(\mathcal{K}^{a_{1}}_{\,\,\,\,ea}
\mathcal{K}^{ea_{2}}_{\,\,\,\,\,\,\,b}\mathcal{K}^{a_{3}}_{\,\,\,\,fc}
\mathcal{K}^{fa_{4}}_{\,\,\,\,\,\,\,d}
\nn\\
&& \ \ \ \ \ \ \ \ \ \ \ \ \ \ \ \
-2\mathcal{K}^{a_{1}\!a_{2}}_{\,\,\,\,\,\,\,\,\,\,a}\mathcal{K}^{a_{3}}_{
\,\,\,\,\,eb}
\mathcal{K}^{e}_{\,\,fc}\mathcal{K}^{fa_{4}}_{\,\,\,\,\,\,\,\,d}
\nn\\
&& \ \ \ \ \ \ \ \ \ \ \ \ \ \ \ \
+2\mathcal{K}^{a_{1}\!a_{2}}_{\,\,\,\,\,\,\,\,\,\,a}\mathcal{K}^{a_{3}}_{
\,\,\,\,\,eb}
\mathcal{K}^{ea_{4}}_{\,\,\,\,\,\,\,f}\mathcal{K}^{f}_{\,\,\,cd}
\nn\\
&& \ \ \ \ \ \ \ \ \ \ \ \ \ \ \ \
+2\mathcal{K}^{a_{1}\!a_{2}}_{\,\,\,\,\,\,\,\,\,\,a}\mathcal{K}^{a_{3}}_{
\,\,\,\,\,eb}
\mathcal{K}^{ea_{4}}_{\,\,\,\,\,\,\,c,d})\delta^{\,a\,b\,c\,d}_{a_{1}a_{2}a_{
3}a_{4}},
\label{GBtelaction}
\end{eqnarray}
where now
\begin{eqnarray}
&&\!\!\!\!\!\!\!\!\!
T_G=(\mathcal{K}^{a_{1}}_{\,\,\,\,ea}\mathcal{K}^{ea_{2}}_{\,\,\,\,\,\,\,b}
\mathcal{K}^{a_{3}}_{\,\,\,\,fc}\mathcal{K}^{fa_{4}}_{\,\,\,\,\,\,\,d}
-2\mathcal{K}^{a_{1}\!a_{2}}_{\,\,\,\,\,\,\,\,\,\,a}\mathcal{K}^{a_{3}}_{
\,\,\,\,\,eb}\mathcal{K}^{e}_{\,\,fc}\mathcal{K}^{fa_{4}}_{\,\,\,\,\,\,\,\,d}
\nn\\
&& \ \ \ \ \,+2\mathcal{K}^{a_{1}\!a_{2}}_{\,\,\,\,\,\,\,\,\,\,a}\mathcal{K}^{a_{3}}_{
\,\,\,\,\,eb}\mathcal{K}^{ea_{4}}_{\,\,\,\,\,\,\,f}\mathcal{K}^{f}_{\,\,\,cd}
\nn\\
&& \ \ \ \ \,+2\mathcal{K}^{a_{1}\!a_{2}}_{\,\,\,\,\,\,\,\,\,\,a}\mathcal{K}^{a_{3}}_{
\,\,\,\,\,eb}\mathcal{K}^{ea_{4}}_{\,\,\,\,\,\,\,c,d})\delta^{\,a\,b\,c\,d}_{a_{1}a_{2}a_{3}a_{4}}\,.
\label{TG}
\end{eqnarray}
The action $S_{tel}^{(2)}$ is a functional of $e^{a}_{\,\,\,\mu}$,
namely $S_{tel}^{(2)}[e^{a}_{\,\,\,\mu}]$, and
although $T_G$ in (\ref{TG}) contains $e^{a}_{\,\,\,\mu,\nu\lambda}$, the
arising equations of motion
contain only $e^{a}_{\,\,\,\mu,\nu\lambda}$ and not higher derivatives, as
expected. The quantity $T_G$ in (\ref{TG})
is a diffeomorphism invariant containing quartic scalars of the torsion (or
contorsion) tensor. However, Lorentz invariance is lost since
preferred autoparallel orthonormal frames have be chosen. As in Einstein gravity,
this is not a deficit, it is a sort of analogue of gauge
fixing in gauge theories. In four dimension, as the general $T_{G}$ of
equation (\ref{poklbn}) is a topological invariant, here $T_{G}$ of equation
(\ref{TG}) is also a topological
invariant constructed out of torsion. This is due to the fact that $T_G$
differs from the  Gauss-Bonnet term, which is topological in four
dimensions, only by a total derivative. Note that the normalization of the actions $S_{Tel}^{(2)}$,
$S_{tel}^{(2)}$ has been defined such that $S_{Tel}^{(2)}=S_{tel}^{(2)}=S_{GB}$.
In case of Einstein-Gauss-Bonnet theory the total action is
$S_{EGB}=S_{EH}+\alpha S_{GB}=S_{tel}^{(1)}+\alpha S_{tel}^{(2)}$, with
$\alpha$ the relevant coupling.

\section{$F(T,T_G)$ gravity and equations of motion}
\label{EF}

In the previous section, we constructed a new quartic-torsion invariant
$T_G$, arising from the Teleparallel Equivalent of Gauss-Bonnet
gravity. Therefore, in analogue with the $F(T)$ gravitational modifications,
we can formulate new modified gravity theories in arbitrary dimensions by
considering general functions $F(T_G)$ in the action. Obviously, since $T_G$
is quartic in torsion, $F(T_G)$ cannot arise from any $F(T)$.
Supplementing the proposed class of modifications with the usual $F(T)$ term,
the total modified gravitational action takes the form
\begin{equation}
S=\frac{1}{2\kappa_D^{2}}\!\int d^{D}\!x\,e\,F(T,T_G)\,,
\label{fGBtelactionD1}
\end{equation}
which is clearly different from $F(R,G)$ gravity \cite{Nojiri:2005jg,DeFelice:2008wz,Davis:2007ta}
(for other constructions of actions including torsion see \cite{zanelli,muller-hoissen}). Obviously,
the usual Einstein-Gauss-Bonnet theory arises in the special case $F(T,T_G)=-T+\alpha T_G$ (with $\alpha$
the Gauss-Bonnet coupling), while TEGR (that is GR) is obtained for $F(T,T_G)=-T$.

In the following, we will extract the equations of motion of $F(T,T_{G})$
gravity by varying the action (\ref{fGBtelactionD1}).
Variation with respect to the vielbein gives
\begin{equation}
2\kappa_{D}^{2}\delta_{e}S=\int\!d^{D}\!x\,\big(eF_{T}\delta_{e}T+eF_{T_{G}}\delta_{e}T_{G}+F\delta e\big)\,,
\label{kwo}
\end{equation}
where $F_{T}=\partial F/\partial T$, $F_{T_{G}}=\partial F/\partial T_{G}$.
Since the variation of $\delta_{e}T_{G}$ is very complicated, we find it more
convenient to make the variations
$\delta_{e}\mathcal{T}_{G}$ and $\delta_{e}\mathcal{T}$ using forms. In
particular, we have
\begin{eqnarray}
2\kappa_{D}^{2}\delta_{e}S&\!\!=\!\!&\int\big(F_{T}\delta_{e}\mathcal{T}+F_{T_{G}}\delta_{e}\mathcal{T}_{G}\big)\nn\\
&\,\,+\!\!&\int \!d^{D}\!x\big(F-TF_{T}-T_{G}F_{T_{G}}\big)\delta e.
\label{njs}
\end{eqnarray}

Let $i_{v}\varphi$ denote the inner derivative of a $p$-form
$\varphi=\frac{1}{p!}\varphi_{a_{1}...a_{p}}e^{a_{1}}
\wedge...\wedge e^{a_{p}}$ with respect to the vector field $v=v^{a}e_{a}$, i.e. for any $p-1$ vector fields
$v_{1},...,v_{p-1}$, it holds
$(i_{v}\varphi)(v_{1},...,v_{p-1})=\varphi(v,v_{1},...,v_{p-1})$.
We are interested in combining this definition with variations. An immediate property is
\begin{equation}
i_{e_{a}}\delta e^{b}+i_{\delta e_{a}}e^{b}=0\,,
\label{ideltae}
\end{equation}
which arises from the equations $\delta e^{a}=e_{b}^{\,\,\,\mu}\delta e^{a}_{\,\,\,\mu}e^{b}$\,,
$\delta e_{a}=e^{b}_{\,\,\,\mu}\delta e_{a}^{\,\,\,\mu}e_{b}$\,,
$i_{e_{a}}\delta e^{b}=e_{a}^{\,\,\,\mu}\delta e^{b}_{\,\,\,\mu}$ and
$i_{\delta e_{a}}e^{b}=e^{b}_{\,\,\,\mu}\delta e_{a}^{\,\,\,\mu}$. Using the definition (\ref{jil}) of the torsion,
equation (\ref{ideltae}), the linearity of $i_{v}\varphi$ in both
$v,\varphi$, and the relations
$i_{v}d+di_{v}=\pounds_{v}$, $i_{v}(\varphi\wedge\psi)=i_{v}\varphi\wedge \psi+(-1)^{p}\varphi\wedge i_{v}\psi$,
we can find
\begin{equation}
\delta_{e}(i_{e_{a}}T^{b})=\pounds_{e_{a}}\delta e^{b}+\pounds_{\delta e_{a}} e^{b}+i_{e_{a}}\omega^{b}_{\,\,\,c}
\wedge\delta e^{c}+i_{\delta e_{a}}\omega^{b}_{\,\,\,c}\wedge e^{c},
\label{whj}
\end{equation}
where $\pounds$ denotes the Lie derivative.

The use of Lie derivative proves
very convenient for the variation procedure. In particular, we use the
identity $v(\alpha(w))=(\pounds_{v}\alpha)(w)+
\alpha(\pounds_{v}w)$, where $\alpha$ is 1-form and $v,w$ are vector fields,
once for $v=\delta e_{a}$, $w=e_{c}$,
$\alpha=e^{b}$ to find $(\pounds_{\delta e_{a}}e^{b})(e_{c})=e^{b}(\pounds_{e_{c}}\delta e_{a})$, and once
for $v=e_{c}$, $w=\delta e_{a}$, $\alpha=e^{b}$ to find $e^{b}(\pounds_{e_{c}}\delta e_{a})=
e_{c}(e^{b}(\delta e_{a}))-(\pounds_{e_{c}}e^{b})(\delta e_{a})$. Therefore, we obtain
\begin{equation}
\pounds_{\delta e_{a}}e^{b}=\pounds_{e_{c}}(e^{b}(\delta e_{a}))e^{c}+C^{b}_{\,\,\,cd}e^{d}(\delta e_{a})e^{c}\,.
\label{drg}
\end{equation}
Thus, the quantity appearing in (\ref{whj}) becomes
\begin{eqnarray}
\delta_{e}(i_{e_{a}}T^{b})=&&\!\!\!\!\!\pounds_{e_{a}}\delta e^{b}+\pounds_{e_{c}}(e^{b}(\delta e_{a}))e^{c}
+C^{b}_{\,\,\,cd}e^{d}(\delta e_{a})e^{c}\nn\\
&&\!\!\!\!+\omega^{b}_{\,\,\,ca}\delta e^{c}+\omega^{b}_{\,\,\,cd}e^{d}(\delta e_{a})e^{c}\,.
\label{kos}
\end{eqnarray}

Additionally, we also need to evaluate the quantity
$\delta_{e}(i_{e_{a}}i_{e_{b}}T^{c})$. Using the definition (\ref{jil})
of the torsion, equation (\ref{ideltae}), the linearity of $i_{v}\varphi$ in both $v,\varphi$, equations
$i_{v}f=0$ ($f$ 0-form),
$i_{v}(\varphi\wedge\psi)=i_{v}\varphi\wedge \psi+(-1)^{p}\varphi\wedge i_{v}\psi$, and the relations
$i_{v}d+di_{v}=\pounds_{v}$, $\pounds_{v}i_{w}-i_{w}\pounds_{v}=i_{[v,w]}$ to transfer the operators $d,\pounds$
on the left, we can find
\begin{eqnarray}
\!\!\!\!\!\!\!\delta_{e}(i_{e_{a}}i_{e_{b}}T^{c})&\!\!=\!\!&i_{[e_{a},e_{b}]}\delta e^{c}+i_{[e_{a},
\delta e_{b}]}e^{c}-i_{[e_{b},\delta e_{a}]}e^{c}\nn\\
&&\!\!
+(i_{e_{b}}\omega^{c}_{\,\,\,d})( i_{e_{a}}\delta e^{d})-(i_{e_{a}}\omega^{c}_{\,\,\,d})( i_{e_{b}}\delta e^{d})\nn\\
&&\!\!
+2\omega^{c}_{\,\,\,[ad]}e^{d}(\delta e_{b})-2\omega^{c}_{\,\,\,[bd]}e^{d}(\delta e_{a})\,,
\label{dfj}
\end{eqnarray}
where the (anti)symmetrization symbol contains the factor $1/2$.
Applying the identity $v(\alpha(w))=(\pounds_{v}\alpha)(w)+\alpha(\pounds_{v}w)$ for $v=e_{a}$, $w=\delta e_{b}$,
$\alpha=e^{c}$, and since $i_{[e_{a},\delta e_{b}]}e^{c}=e^{c}(\pounds_{e_{a}}\delta e_{b})$, we find
\begin{equation}
i_{[e_{a},\delta e_{b}]}e^{c}=\pounds_{e_{a}}(e^{c}(\delta e_{b}))+C^{c}_{\,\,\,ad}e^{d}(\delta e_{b})\,.
\label{dne}
\end{equation}
Finally, using (\ref{ideltae}), we acquire
\begin{eqnarray}
\!\!\!\delta_{e}(i_{e_{a}}i_{e_{b}}T^{c})&\!\!\!=\!\!\!&
\pounds_{e_{a}}(e^{c}(\delta e_{b}))-\pounds_{e_{b}}(e^{c}(\delta e_{a}))\nn\\
&&\!\!\!
+C^{c}_{\,\,\,ad}e^{d}(\delta e_{b})-C^{c}_{\,\,\,bd}e^{d}(\delta e_{a})-C^{d}_{\,\,\,ab}e^{c}(\delta e_{d})\nn\\
&&\!\!\!
+\omega^{c}_{\,\,\,ad}e^{d}(\delta e_{b})-\omega^{c}_{\,\,\,bd}e^{d}(\delta e_{a})\,.
\label{sja}
\end{eqnarray}

Now, the contorsion 1-form can be written as
\begin{equation}
2\mathcal{K}_{ab}=i_{e_{a}}T_{b}-i_{e_{b}}T_{a}-(i_{e_{a}}i_{e_{b}}T_{c})e^{c}\,,
\label{gso}
\end{equation}
therefore we obtain
\begin{equation}
2\delta_{e}\mathcal{K}_{ab}=\delta_{e}i_{e_{a}}T_{b}-\delta_{e}i_{e_{b}}T_{a}-\delta_{e}(i_{e_{a}}i_{e_{b}}T_{c})e^{c}
-T_{cba}\delta e^{c}\,.
\label{hdj}
\end{equation}
Using (\ref{kos}) and (\ref{sja}) we get
\begin{eqnarray}
2\delta_{e}\mathcal{K}_{ab}&\!\!\!=\!\!\!&
\pounds_{\!e_{a}}\!\delta \underline{e}_{b}\!-\!\pounds_{\!e_{b}}\!\delta \underline{e}_{a}
\!+\!\pounds_{\!e_{c}}\!(i_{e_{b}}\delta\underline{e}_{a})e^{c}
\!-\!\pounds_{\!e_{c}}\!(i_{e_{a}}\delta\underline{e}_{b})e^{c}\nn\\
&&\!\!\!\!\!+\pounds_{\!e_{a}}\!(i_{e_{b}}\delta\underline{e}_{c})e^{c}
\!-\!\pounds_{\!e_{b}}\!(i_{e_{a}}\delta\underline{e}_{c})e^{c}
\!-\!C^{d}_{\,\,\,ab}(i_{e_{d}}\delta\underline{e}_{c})e^{c}\nn\\
&&\!\!\!\!\!
+2C_{(ac)d}(i_{e_{b}}\delta e^{d})e^{c}\!-\!2C_{(bc)d}(i_{e_{a}}\delta e^{d})e^{c}\!+\!T_{cab}\delta e^{c}
\nn\\
&&\!\!\!\!\!\!+2\omega_{c[ab]}\delta e^{c}\,,
\label{kza}
\end{eqnarray}
where $\underline{e}_{a}=\eta_{ab}e^{b}$ are 1-forms and
$\underline{e}_{b}(\delta e_{a})=-i_{e_{a}}
\delta \underline{e}_{b}$.

Varying $\mathcal{T},\mathcal{T}_{G}$ from (\ref{tosrionscalarform}) and
(\ref{pokl}), and
due to the fact that $i_{e_{a}}\delta e^{a}=-e^{a}_{\,\,\,\mu}
\delta e_{a}^{\,\,\,\mu}=\frac{\delta e}{e}$, the variation (\ref{njs}) of the action becomes
\begin{eqnarray}
\!\!\!\!\!\!\!\!
2\kappa_{D}^{2}\delta_{e}S\!&\!=\!&\!\!\!\int\!\big(2\delta_{e}\mathcal{K}_{ab}\wedge H^{ab}+
h_{a}\wedge \delta e^{a}\big)\nn\\
&\!\!+\!\!\!\!&\!\!\int\!\big(F\!-\!TF_{T}\!-\!T_{G}F_{T_{G}}\big)(i_{e_{a}}\delta e^{a})e^{1}\!\wedge...\wedge\!e^{D},
\label{kso}
\end{eqnarray}
where
\begin{eqnarray}
&&\!\!\!\!\!\!\!\!\!\!\!\!\!\!\!\!\!\!\!
H^{ab}\!=\!\frac{F_{T}}{(D\!-\!2)!}\epsilon^{a}_{\,\,\,a_{1}...a_{D-1}}\mathcal{K}^{ba_{1}}
e^{a_{2}}...e^{a_{D-1}}\nn\\
&&\!\!\!
+\frac{F_{T_{G}}}{(D\!-\!4)!}\big(2\epsilon^{a}_{\,\,\,a_{1}...a_{D-1}}
\mathcal{K}^{ba_{1}}
\mathcal{K}^{a_{2}}_{\,\,\,\,\,c}\mathcal{K}^{ca_{3}}e^{a_{4}}... e^{a_{D-1}}\nn\\
&&\,\,\,\,\,\,\,\,\,\,\,\,\,\,\,\,\,\,\,\,\,\,\,\,\,\,
+\epsilon_{a_{1}...a_{D}}\mathcal{K}^{aa_{1}}\mathcal{K}^{ba_{2}}\mathcal{K}^{a_{3}a_{4}}e^{a_{5}}...e^{a_{D}}\nn\\
&&\,\,\,\,\,\,\,\,\,\,\,\,\,\,\,\,\,\,\,\,\,\,\,\,\,\,
-\epsilon^{ab}_{\,\,\,\,\,\,a_{1}...a_{D-2}}\mathcal{K}^{a_{1}}_{\,\,\,\,\,c}\mathcal{K}^{c}_{\,\,\,d}
\mathcal{K}^{da_{2}}e^{a_{3}}...e^{a_{D-2}}\nn\\
&&\,\,\,\,\,\,\,\,\,\,\,\,\,\,\,\,\,\,\,\,\,\,\,\,\,\,
+\epsilon^{ab}_{\,\,\,\,\,\,a_{1}...a_{D-2}}D\mathcal{K}^{a_{1}}_{\,\,\,\,\,c}\mathcal{K}^{ca_{2}}
e^{a_{3}}...e^{a_{D-2}}\nn\\
&&\,\,\,\,\,\,\,\,\,\,\,\,\,\,\,\,\,\,\,\,\,\,\,\,\,\,
+\epsilon^{a}_{\,\,\,a_{1}...a_{D-1}}D\mathcal{K}^{ba_{1}}
\mathcal{K}^{a_{2}a_{3}}e^{a_{4}}... e^{a_{D-1}}\big)\nn\\
&&\!\!\!\!\!\!-\frac{1}{(D\!-\!4)!}\epsilon^{a}_{\,\,\,a_{1}...a_{D-1}}
D(F_{T_{G}}\mathcal{K}^{ba_{1}}\mathcal{K}^{a_{2}a_{3}}e^{a_{4}}...e^{a_{D-1}})
\label{asb}
\end{eqnarray}
and
\begin{eqnarray}
&&\!\!\!\!\!\!\!\!\!\!\!\!
h_{a}\!=\!\frac{F_{T}}{(D\!-\!3)!}\epsilon_{a_{1}...a_{D-1}a}\mathcal{K}^{a_{1}}_{\,\,\,\,c}
\mathcal{K}^{ca_{2}} e^{a_{3}}...e^{a_{D-1}}\nn\\
&&+\frac{F_{T_{G}}}{(D\!-\!5)!}
\epsilon_{a_{1}...a_{D-1}a}\big(\mathcal{K}^{a_{1}}_{\,\,\,\,c}
\mathcal{K}^{ca_{2}}\mathcal{K}^{a_{3}}_{\,\,\,\,\,d}\mathcal{K}^{da_{4}}\nn\\
&&\,\,\,\,\,\,\,\,\,\,\,\,\,\,\,\,\,\,\,\,\,\,\,\,\,\,\,\,\,\,\,\,\,\,\,\,\,\,\,\,\,\,\,\,\,\,\,\,\,\,\,\,
-2\mathcal{K}^{a_{1}a_{2}}\mathcal{K}^{a_{3}}_{\,\,\,\,\,c}\mathcal{K}^{c}_{\,\,\,d}\mathcal{K}^{da_{4}}\nn\\
&&\,\,\,\,\,\,\,\,\,\,\,\,\,\,\,\,\,\,\,\,\,\,\,\,\,\,\,\,\,\,\,\,\,\,\,\,
+2\mathcal{K}^{a_{1}a_{2}}D\mathcal{K}^{a_{3}}_{\,\,\,\,\,c}\mathcal{K}^{ca_{4}}\big) e^{a_{5}}... e^{a_{D-1}}\,.
\label{sdo}
\end{eqnarray}
The quantities
$H^{ab},h_{a}$ are $D-1$ forms and the $\wedge$ symbols between $\mathcal{K}^{ab}$ and $e^{a}$ are omitted for safety
of space. Moreover, boundary terms have been omitted too. The above
relations hold for $D>4$, while for $D=4$ all terms exist too, apart from the
term containing $(D-5)!$ in $h_{a}$, which is absent.

Now, we plague   expression (\ref{kza}) in the variation (\ref{kso}), and
after use of the identity
$\pounds_{e_{a}}=i_{e_{a}}d+di_{e_{a}}$ and the obvious equations
$i_{e_{b}}(\pounds_{e_{c}}(e^{c}\wedge H^{[ab]})\wedge \delta\underline{e}_{a})=0$,
$i_{e_{b}}(\pounds_{e_{a}}(e^{c}\wedge H^{[ab]})\wedge \delta\underline{e}_{c})=0$,
$i_{e_{d}}(C^{d}_{\,\,\,ab}e^{c}\wedge H^{ab}\wedge\delta\underline{e}_{c})=0$,
$i_{e_{b}}(C_{(ac)d}e^{c}\wedge H^{ab}\wedge\delta e^{d})=0$, we obtain (omitting the boundary terms)
\begin{eqnarray}
&&\!\!\!\!\!\!2\kappa_{D}^{2}\delta_{e}S=\!\!\int\!\delta\underline{e}_{a} \Big[2\pounds_{e_{b}}H^{[ab]}
\!-\!2i_{e_{b}}\pounds_{e_{c}}(e^{c}H^{[ab]}\!+\!e^{a}H^{[cb]})\nn\\
&&\!\!\!\!\!\!\!-C^{d}_{\,\,\,cb}i_{e_{d}}\!(e^{a}H^{cb})\!+\!4C_{(dc)}^{\,\,\,\,\,\,\,\,a}i_{e_{b}}\!(e^{c}
H^{[db]})\!+\!(T^{a}_{\,\,\,bc}\!+\!2\omega^{a}_{\,\,\,[bc]})H^{bc}\nn\\
&&\,\,\,\,\,\,\,\,\,\,\,\,\,\,\,\,\,\,\,\,\,\,\,\,\,\,\,\,\,\,\,\,\,\,
-(-1)^{D}h^{a}\!+\!(F\!-\!TF_{T}\!-\!T_{G}F_{T_{G}})\vartheta^{a}\Big]\,,
\label{shj}
\end{eqnarray}
where $\vartheta_{a}=i_{e_{a}}(e^{1}\wedge...\wedge e^{D})$. Thus, setting
$\delta_{e}S=0$ we get {\textit{the equations of motion for $F(T,T_{G})$ gravity}}
\begin{eqnarray}
&&\!\!\!\!\!\!2\pounds_{e_{b}}H^{[ab]}
\!-\!2i_{e_{b}}\pounds_{e_{c}}(e^{c}H^{[ab]}\!+\!e^{a}H^{[cb]})\!-\!C^{d}_{\,\,\,cb}\,i_{e_{d}}\!(e^{a}H^{cb})\nn\\
&&+4C_{(dc)}^{\,\,\,\,\,\,\,\,a}\,i_{e_{b}}\!(e^{c}
H^{[db]})\!+\!(T^{a}_{\,\,\,bc}\!+\!2\omega^{a}_{\,\,\,[bc]})H^{bc}\!-\!(-1)^{D}h^{a}\nn\\
&&\,\,\,\,\,\,\,\,\,\,\,\,\,\,\,\,\,\,\,\,\,\,\,\,\,\,\,\,\,\,\,\,\,\,\,\,\,\,\,\,\,\,\,\,\,\,\,\,
+(F\!-\!TF_{T}\!-\!T_{G}F_{T_{G}})\vartheta^{a}=0\,.
\label{jkk}
\end{eqnarray}
The set $\vartheta_{a}$ forms a basis in the subspace of $D-1$ forms, therefore $H^{ab},h^{a}$ can be expressed in
components as
\begin{equation}
H^{ab}=H^{abc}\vartheta_{c}\,\,\,\,\,\,,\,\,\,\,\,\,h^{a}=h^{ab}\vartheta_{b}\,.
\label{khq}
\end{equation}
Hence, the equations of motion (\ref{jkk}) for $F(T,T_{G})$ gravity is
written in components as
\begin{eqnarray}
&&\!\!\!\!\!\!
2(H^{[ac]b}\!+\!H^{[ba]c}\!-\!H^{[cb]a})_{,c}\!+\!2(H^{[ac]b}\!+\!H^{[ba]c}\!-\!H^{[cb]a})C^{d}_{\,\,\,dc}\nn\\
&&\!\!\!\!\!\!\!\!+(2H^{[ac]d}\!+\!H^{dca})C^{b}_{\,\,\,cd}
\!+\!4H^{[db]c}C_{(dc)}^{\,\,\,\,\,\,\,\,a}\!+\!(T^{a}_{\,\,\,cd}\!+\!2\omega^{a}_{\,\,\,[cd]})H^{cdb}\nn\\
&&\,\,\,\,\,\,\,\,\,\,\,\,\,\,\,\,\,\,\,\,\,\,\,-(-1)^{D}h^{ab}\!+\!(F\!-\!TF_{T}\!-\!T_{G}F_{T_{G}})\eta^{ab}=0\,.
\label{gkw}
\end{eqnarray}

Focusing to the most interesting case of four dimensions we can re-write the
expressions for $H^{ab}$ and $h_{a}$ as
\begin{eqnarray}
&&\!\!\!\!\!\!\!\!\!\!\!\!\!\!\!\!
H^{ab}\!=\!\frac{F_{T}}{2}\epsilon^{a}_{\,\,\,cdf}\mathcal{K}^{bc}
e^{d}e^{f}-\epsilon^{a}_{\,\,\,cdf}
D(F_{T_{G}}\mathcal{K}^{bc}\mathcal{K}^{df})\nn\\
&&
+F_{T_{G}}\big(2\epsilon^{a}_{\,\,\,cdf}
\mathcal{K}^{bc}\mathcal{K}^{d}_{\,\,\,q}\mathcal{K}^{qf}
+\epsilon_{cdfq}
\mathcal{K}^{ac}\mathcal{K}^{bd}\mathcal{K}^{fq}\nn\\
&&\,\,\,\,\,\,\,\,\,\,\,\,\,\,\,\,\,\,
-\epsilon^{ab}_{\,\,\,\,\,\,cd}\mathcal{K}^{c}_{\,\,\,f}\mathcal{K}^{f}_{\,\,\,q}
\mathcal{K}^{qd}
+\epsilon^{ab}_{\,\,\,\,\,\,cd}D\mathcal{K}^{c}_{\,\,\,f}\mathcal{K}^{fd}\nn\\
&&\,\,\,\,\,\,\,\,\,\,\,\,\,\,\,\,\,\,
+\epsilon^{ac}_{\,\,\,\,\,\,df}D\mathcal{K}^{b}_{\,\,\,c}
\mathcal{K}^{df}\big)
\label{kju}
\end{eqnarray}
and
\begin{eqnarray}
&&\!\!\!\!\!\!\!\!\!\!\!\!
h_{a}\!=\!-F_{T}\epsilon_{abcd}\mathcal{K}^{b}_{\,\,\,f}
\mathcal{K}^{fc} e^{d}\,.
\label{mkd}
\end{eqnarray}
Since $e^{a}\wedge e^{b}\wedge e^{c}=\epsilon^{abcd}\vartheta_{d}$, $e^{a}\wedge e^{b}=
-\frac{1}{2}\epsilon^{abcd}\vartheta_{cd}$, where
$\vartheta_{ab}=i_{e_{b}}\vartheta_{a}$, the above expressions become
\begin{eqnarray}
&& \!\!\!\!\!\!\!
H^{abc}=F_{T}(\eta^{ac}\mathcal{K}^{bd}_{\,\,\,\,\,\,d}-\mathcal{K}^{bca})+F_{T_{G}}\big[\nn\\
&&\!\!\!\!\!\!\!\epsilon^{cprt}\!\big(\!2\epsilon^{a}_{\,\,\,dkf}\mathcal{K}^{bk}_{\,\,\,\,\,p}
\mathcal{K}^{d}_{\,\,\,qr}
\!\!+\!\epsilon_{qdkf}\mathcal{K}^{ak}_{\,\,\,\,\,p}\mathcal{K}^{bd}_{\,\,\,\,\,\,r}\!\!+\!
\epsilon^{ab}_{\,\,\,\,\,\,kf}\mathcal{K}^{k}_{\,\,\,dp}\mathcal{K}^{d}_{\,\,\,qr}\!\big)\!
\mathcal{K}^{qf}_{\,\,\,\,\,\,t}\nn\\
&&\!\!\!\!\!\!\!+\epsilon^{cprt}\epsilon^{ab}_{\,\,\,\,\,\,kd}\mathcal{K}^{fd}_{\,\,\,\,\,\,p}
\big(\mathcal{K}^{k}_{\,\,fr,t}\!-\!\frac{1}{2}\mathcal{K}^{k}_{\,\,fq}C^{q}_{\,\,\,tr}
\!\!+\!\omega^{k}_{\,\,\,qt}\mathcal{K}^{q}_{\,\,fr}\!\!+\!\omega^{q}_{\,\,fr}\mathcal{K}^{k}_{\,\,qt}\big)\nn\\
&&\!\!\!\!\!\!\!+\epsilon^{cprt}\epsilon^{ak}_{\,\,\,\,\,\,df}\mathcal{K}^{df}_{\,\,\,\,p}
\big(\mathcal{K}^{b}_{\,\,kr,t}\!-\!\frac{1}{2}\mathcal{K}^{b}_{\,\,kq}C^{q}_{\,\,\,tr}
\!\!+\!\omega^{b}_{\,\,\,qt}\mathcal{K}^{q}_{\,\,kr}\!\!+\!\omega^{q}_{\,\,\,kr}\mathcal{K}^{b}_{\,\,qt}\big)\big]
\nn\\
&&\!\!\!\!\!\!\!+F_{T_{G}}\epsilon^{cprt}\epsilon^{a}_{\,\,\,kdf}\Big[
\frac{1}{F_{T_{G}}}\big(F_{T_{G}}\mathcal{K}^{bk}_{\,\,\,\,\,p}
\mathcal{K}^{df}_{\,\,\,\,\,r}\big)_{,t}\!+\!
C^{q}_{\,\,\,pt}\mathcal{K}^{bk}_{\,\,\,\,\,[q}\mathcal{K}^{df}_{\,\,\,\,\,r]}
\nn\\
&&\!\!\!\!\!\!\!
+(\omega^{b}_{\,\,\,qp}\mathcal{K}^{qk}_{\,\,\,\,\,\,r}\!+\!\omega^{k}_{\,\,\,qp}\mathcal{K}^{bq}_{\,\,\,\,\,\,r})
\mathcal{K}^{df}_{\,\,\,\,\,t}
\!+\!(\omega^{d}_{\,\,\,qp}\mathcal{K}^{qf}_{\,\,\,\,\,\,t}\!+\!\omega^{f}_{\,\,\,qp}\mathcal{K}^{dq}_{\,\,\,\,\,\,t})
\mathcal{K}^{bk}_{\,\,\,\,\,r}\Big]\label{kas}\nn\\
\end{eqnarray}
\begin{equation}
h^{ab}=F_{T}\epsilon^{a}_{\,\,\,kcd}\epsilon^{bpqd}\mathcal{K}^{k}_{\,\,\,fp}\mathcal{K}^{fc}_{\,\,\,\,\,\,q}\,.
\label{ksl}
\end{equation}

Choosing additionally the Weitzenb{\"{o}}ck connection
$\omega^{a}_{\,\,\,bc}=0$, for $D=4$ we finally obtain
\begin{eqnarray}
&&\!\!\!\!\!\!
2(H^{[ac]b}\!+\!H^{[ba]c}\!-\!H^{[cb]a})_{,c}\!+\!2(H^{[ac]b}\!+\!H^{[ba]c}\!-\!H^{[cb]a})C^{d}_{\,\,\,dc}\nn\\
&&\!\!\!\!\!\!\!\!+(2H^{[ac]d}\!+\!H^{dca})C^{b}_{\,\,\,cd}
\!+\!4H^{[db]c}C_{(dc)}^{\,\,\,\,\,\,\,\,a}\!+\!T^{a}_{\,\,\,cd}H^{cdb}-h^{ab}\nn\\
&&\,\,\,\,\,\,\,\,\,\,\,\,\,\,\,\,\,\,\,\,\,\,\,\,\,\,\,\,\,\,\,\,\,\,\,\,\,\,\,\,\,\,\,\,\,\,\,\,\,\,
+(F\!-\!TF_{T}\!-\!T_{G}F_{T_{G}})\eta^{ab}=0\,,
\label{kal}
\end{eqnarray}
where
\begin{eqnarray}
&& \!\!\!\!\!\!\!
H^{abc}=F_{T}(\eta^{ac}\mathcal{K}^{bd}_{\,\,\,\,\,\,d}-\mathcal{K}^{bca})+F_{T_{G}}\big[\nn\\
&&\!\!\!\!\!\!\!\epsilon^{cprt}\!\big(\!2\epsilon^{a}_{\,\,\,dkf}\mathcal{K}^{bk}_{\,\,\,\,\,p}
\mathcal{K}^{d}_{\,\,\,qr}
\!\!+\!\epsilon_{qdkf}\mathcal{K}^{ak}_{\,\,\,\,\,p}\mathcal{K}^{bd}_{\,\,\,\,\,\,r}\!\!+\!
\epsilon^{ab}_{\,\,\,\,\,\,kf}\mathcal{K}^{k}_{\,\,\,dp}\mathcal{K}^{d}_{\,\,\,qr}\!\big)\!
\mathcal{K}^{qf}_{\,\,\,\,\,\,t}\nn\\
&&\,\,\,\,\,\,\,\,\,\,\,\,\,\,\,\,\,\,\,\,\,\,\,\,\,\,\,\,\,\,\,\,\,\,\,\,\,\,\,\,\,\,\,\,
+\epsilon^{cprt}\epsilon^{ab}_{\,\,\,\,\,\,kd}\mathcal{K}^{fd}_{\,\,\,\,\,\,p}
\big(\mathcal{K}^{k}_{\,\,fr,t}\!-\!\frac{1}{2}\mathcal{K}^{k}_{\,\,fq}C^{q}_{\,\,\,tr}\big)\nn\\
&&\,\,\,\,\,\,\,\,\,\,\,\,\,\,\,\,\,\,\,\,\,\,\,\,\,\,\,\,\,\,\,\,\,\,\,\,\,\,\,\,\,\,\,
+\epsilon^{cprt}\epsilon^{ak}_{\,\,\,\,\,\,df}\mathcal{K}^{df}_{\,\,\,\,p}
\big(\mathcal{K}^{b}_{\,\,kr,t}\!-\!\frac{1}{2}\mathcal{K}^{b}_{\,\,kq}C^{q}_{\,\,\,tr}\big)\big]
\nn\\
&&\!\!+\epsilon^{cprt}\epsilon^{a}_{\,\,\,kdf}\Big[
\big(F_{T_{G}}\mathcal{K}^{bk}_{\,\,\,\,\,p}
\mathcal{K}^{df}_{\,\,\,\,\,r}\big)_{,t}\!+\!
F_{T_{G}}C^{q}_{\,\,\,pt}\mathcal{K}^{bk}_{\,\,\,\,\,[q}\mathcal{K}^{df}_{\,\,\,\,\,r]}\Big]\label{ksi}
\end{eqnarray}
and
\begin{equation}
h^{ab}=F_{T}\epsilon^{a}_{\,\,\,kcd}\epsilon^{bpqd}\mathcal{K}^{k}_{\,\,\,fp}\mathcal{K}^{fc}_{\,\,\,\,\,\,q}\,.
\label{hfp}
\end{equation}

Equations (\ref{kal}) are the equations of motion for $F(T,T_G)$ gravity in
four dimensions, for a general vielbein (or equivalently for a general
metric) choice. For specific cases, such as the homogeneous and
isotropic Friedmann-Robertson-Walker and the spherically symmetric
geometries, the above equations are significantly simplified. Thus, one can
straightforwardly investigate the application of $F(T,T_G)$ gravity in a
cosmological framework. Since this study lies beyond the scope of the
present work, it is left for a separate project \cite{uscosmology}.

\section{Conclusions} \label{Conclusions}

Inspired by the teleparallel formulation of General Relativity, whose
Lagrangian is the torsion invariant $T$, we have constructed the teleparallel
equivalent of Gauss-Bonnet gravity in arbitrary dimensions. Implementing the
teleparallel condition, but without imposing the Weitzenb{\"{o}}ck
connection, we have extracted the torsion invariant $T_G$, equivalent (up to boundary
terms) to the Gauss-Bonnet term $G$. $T_{G}$ is made out of the vielbein $e^{a}$ and
the connection $\omega^{a}_{\,\,\,b}$, it contains quartic powers of the torsion tensor,
and it is diffeomorphism and Lorentz invariant. In four dimensions it reduces to a topological
invariant, as expected. Imposing the Weitzenb{\"{o}}ck connection, a simpler
form for $T_G$ arises containing only the vielbein. This allows us to define a new
class of modified gravity theories based on $F(T,T_G)$, which is not spanned by the
class of $F(T)$ theories. Moreover, it is also distinct from the $F(R,G)$ class.
Hence, $F(T,T_G)$ theory is a novel class of modified gravity.
Finally, varying the action with respect to the vielbein, we extracted the
equations of motion for a general vielbein (metric) choice. Since
$F(T,T_G)$ gravity is a new modified gravitational theory, it would be
interesting to study its cosmological applications, and this is performed in
a separate publication \cite{uscosmology}.

\begin{acknowledgments}
The research of ENS is implemented within the framework of the
Operational Program ``Education and Lifelong Learning'' (Actions
Beneficiary: General Secretariat for Research and Technology), and
is co-financed by the European Social Fund (ESF) and the Greek State.
\end{acknowledgments}



\begin{thebibliography}{99}


\bibitem{Weinberg:2008}
  S. Weinberg, \emph{Cosmology},
  Oxford University Press Inc., New York (2008).


\bibitem{ein28}
A. Einstein 1928, Sitz. Preuss. Akad. Wiss. p. 217; ibid p. 224;
  A.~Unzicker and T.~Case
     [arXiv:physics/0503046].


\bibitem{DeFelice:2010aj}
  A.~De Felice and S.~Tsujikawa,
  Living Rev.\ Rel.\  {\bf 13}, 3 (2010).


\bibitem{Nojiri:2010wj}
  S.~'i.~Nojiri and S.~D.~Odintsov,
  Phys.\ Rept.\  {\bf 505}, 59 (2011).


\bibitem{Wheeler:1985nh} D. Boulware and S. Deser, Phys. Rev. Lett. {\bf 55},
 2656 (1985);
J. T. Wheeler, Nucl. Phys. B {\bf 268}, 737 (1986)


\bibitem{Antoniadis:1993jc}
  I.~Antoniadis, J.~Rizos and K.~Tamvakis,
  Nucl.\ Phys.\ B {\bf 415}, 497 (1994);
  P.~Kanti, J.~Rizos and K.~Tamvakis,
  Phys.\ Rev.\ D {\bf 59}, 083512 (1999);
  N.~E.~Mavromatos and J.~Rizos,
  Phys.\ Rev.\ D {\bf 62}, 124004 (2000);
  S.~'i.~Nojiri, S.~D.~Odintsov and M.~Sasaki,
  Phys.\ Rev.\ D {\bf 71}, 123509 (2005).


\bibitem{Nojiri:2005jg}
  S.~'i.~Nojiri and S.~D.~Odintsov,
  Phys.\ Lett.\ B {\bf 631}, 1 (2005).


\bibitem{DeFelice:2008wz}
  A.~De Felice and S.~Tsujikawa,
  Phys.\ Lett.\ B {\bf 675}, 1 (2009).


\bibitem{Lovelock:1971yv}
  D.~Lovelock,
  J.\ Math.\ Phys.\  {\bf 12}, 498 (1971).


\bibitem{Deruelle:1989fj}
  N.~Deruelle and L.~Farina-Busto,
  Phys.\ Rev.\ D {\bf 41}, 3696 (1990).


\bibitem{Mannheim:1988dj}
  P.~D.~Mannheim and D.~Kazanas,
  Astrophys.\ J.\  {\bf 342}, 635 (1989);
  E.~E.~Flanagan,
  Phys.\ Rev.\ D {\bf 74}, 023002 (2006);
  N.~Deruelle, M.~Sasaki, Y.~Sendouda and A.~Youssef,
  JCAP {\bf 1103}, 040 (2011);
  D.~Grumiller, M.~Irakleidou, I.~Lovrekovic and R.~McNees,
  Phys.\ Rev.\ Lett.\  {\bf 112}, 111102 (2014).


\bibitem{Horava:2008ih}
  P.~Horava,
  JHEP {\bf 0903}, 020 (2009);
  P.~Horava,
  Phys.\ Rev.\  D {\bf 79}, 084008 (2009);
  G.~Calcagni,
  JHEP {\bf 0909}, 112 (2009);
  E.~Kiritsis and G.~Kofinas,
  Nucl.\ Phys.\ B {\bf 821}, 467 (2009);
  E.~N.~Saridakis,
  Eur.\ Phys.\ J.\ C {\bf 67}, 229 (2010).



\bibitem{Ferraro:2006jd}
  R.~Ferraro and F.~Fiorini,
  Phys.\ Rev.\ D {\bf 75}, 084031 (2007);
  R.~Ferraro, F.~Fiorini,
  Phys.\ Rev.\  {\bf D78}, 124019 (2008);
  G.~R.~Bengochea and R.~Ferraro,
  Phys.\ Rev.\ D \textbf{79}, 124019 (2009).


\bibitem{Linder:2010py}
  E.~V.~Linder,
  Phys.\ Rev.\ D \textbf{81}, 127301 (2010).


\bibitem{Myrzakulov:2010vz}
S.~H.~Chen, J.~B.~Dent, S.~Dutta and E.~N.~Saridakis,
Phys.\ Rev.\ D
\textbf{ 83}, 023508 (2011);
J.~B.~Dent, S.~Dutta, E.~N.~Saridakis,
  JCAP {\bf 1101}, 009 (2011);
  T.~P.~Sotiriou, B.~Li and J.~D.~Barrow,
  Phys.\ Rev.\ D {\bf 83}, 104030 (2011);
   Y.~Zhang, H.~Li, Y.~Gong, Z.~-H.~Zhu,
  JCAP {\bf 1107}, 015 (2011);
  Y.~-F.~Cai, S.~-H.~Chen, J.~B.~Dent, S.~Dutta, E.~N.~Saridakis,
  Class.\ Quant.\ Grav.\  {\bf 28}, 2150011 (2011);
  M.~Sharif, S.~Rani,
  Mod.\ Phys.\ Lett.\  {\bf A26}, 1657 (2011);
   S.~Capozziello, V.~F.~Cardone, H.~Farajollahi and A.~Ravanpak,
  Phys.\ Rev.\ D {\bf 84}, 043527 (2011);
  C.~-Q.~Geng, C.~-C.~Lee, E.~N.~Saridakis, Y.~-P.~Wu,
  Phys.\ Lett.\  {\bf B704}, 384 (2011);
    H.~Wei,
  Phys.\ Lett.\ B {\bf 712}, 430 (2012);
  C.~-Q.~Geng, C.~-C.~Lee, E.~N.~Saridakis,
  JCAP {\bf 1201}, 002 (2012);
    Y.~-P.~Wu and C.~-Q.~Geng,
  Phys.\ Rev.\ D {\bf 86}, 104058 (2012);
  C.~G.~Bohmer, T.~Harko and F.~S.~N.~Lobo,
  Phys.\ Rev.\ D {\bf 85}, 044033 (2012);
    K.~Karami and A.~Abdolmaleki,
  JCAP {\bf 1204}, 007 (2012);
   C.~Xu, E.~N.~Saridakis and G.~Leon,
  JCAP {\bf 1207}, 005 (2012);
  K.~Bamba, R.~Myrzakulov, S.~'i.~Nojiri and S.~D.~Odintsov,
Phys. Rev. D {\bf 85}, 104036 (2012);
  N.~Tamanini and C.~G.~Boehmer,
  Phys.\ Rev.\ D {\bf 86}, 044009 (2012);
  M.~E.~Rodrigues, M.~J.~S.~Houndjo, D.~Saez-Gomez and F.~Rahaman,
  Phys.\ Rev.\ D {\bf 86}, 104059 (2012);
  K.~Bamba, J.~de Haro and S.~D.~Odintsov,
  JCAP {\bf 1302}, 008 (2013);
  M.~Jamil, D.~Momeni and R.~Myrzakulov,
  Eur.\ Phys.\ J.\ C {\bf 72}, 2267 (2012);
  J.~-T.~Li, C.~-C.~Lee and C.~-Q.~Geng,
  Eur.\ Phys.\ J.\ C {\bf 73}, 2315 (2013);
  A.~Aviles, A.~Bravetti, S.~Capozziello and O.~Luongo,
  Phys.\ Rev.\ D {\bf 87}, 064025 (2013);
   Y.~C.~Ong, K.~Izumi, J.~M.~Nester and P.~Chen,
 Phys.\ Rev.\ D {\bf 88}, 024019 (2013);
  K.~Bamba, S.~'i.~Nojiri and S.~D.~Odintsov,
  arXiv:1304.6191 [gr-qc];
 G.~Otalora,
 JCAP {\bf 1307}, 044 (2013);
  J.~Amoros, J.~de Haro and S.~D.~Odintsov,
     Phys.\ Rev.\ D {\bf 87}, 104037 (2013);
  G.~Otalora,
  Phys.\ Rev.\ D {\bf 88}, 063505 (2013);
  C.~-Q.~Geng, J.~-A.~Gu and C.~-C.~Lee,
   Phys.\ Rev.\ D {\bf 88}, 024030 (2013);
  M.~E.~Rodrigues, I.~G.~Salako, M.~J.~S.~Houndjo and J.~Tossa,
Int. J. Mod. Phys. D {\bf 23} (2014) 1, 1450004;
   K.~Bamba, S.~Capozziello, M.~De Laurentis, S.~'i.~Nojiri and
D.~Sáez-Gómez,
  Phys.\ Lett.\ B {\bf 727}, 194 (2013);
  K.~Bamba, S.~'i.~Nojiri and S.~D.~Odintsov,
  Phys.\ Lett.\ B {\bf 731}, 257 (2014);
  G.~Otalora,
  arXiv:1402.2256 [gr-qc];
  A.~Paliathanasis, S.~Basilakos, E.~N.~Saridakis, S.~Capozziello,
K.~Atazadeh, F.~Darabi and M.~Tsamparlis,
  arXiv:1402.5935 [gr-qc];
  G.~G.~L.~Nashed,
  arXiv:1403.6937 [gr-qc];
  S.~Chattopadhyay,
  arXiv:1403.8116 [gr-qc];
  G.~Kofinas, G.~Leon and E.~N.~Saridakis,
  Class.\ Quant.\ Grav.\  {\bf 31}, 175011 (2014);
  V.~Fayaz, H.~Hossienkhani, A.~Farmany, M.~Amirabadi and N.~Azimi,
  Astrophys.\ Space Sci.\  {\bf 351}, 299 (2014);
  T.~Harko, F.~S.~N.~Lobo, G.~Otalora and E.~N.~Saridakis,
  Phys.\ Rev.\ D {\bf 89}, 124036 (2014);
  T.~Harko, F.~S.~N.~Lobo, G.~Otalora and E.~N.~Saridakis,
  arXiv:1405.0519 [gr-qc].



\bibitem{Wu:2010mn}
  P.~Wu, H.~W.~Yu,
  Phys.\ Lett.\ \textbf{B693}, 415 (2010);
    G.~R.~Bengochea,
  Phys.\ Lett.\  {\bf B695}, 405 (2011);
  L.~Iorio and E.~N.~Saridakis,
  Mon.\ Not.\ Roy.\ Astron.\ Soc.\  {\bf 427}, 1555 (2012);
  S.~Nesseris, S.~Basilakos, E.~N.~Saridakis and L.~Perivolaropoulos,
  Phys.\ Rev.\ D {\bf 88}, 103010 (2013).


\bibitem{Wang:2011xf}
    T.~Wang,
  Phys.\ Rev.\  {\bf D84}, 024042 (2011);
  R.~-X.~Miao, M.~Li and Y.~-G.~Miao,
  JCAP {\bf 1111}, 033 (2011);
  C.~G.~Boehmer, A.~Mussa and N.~Tamanini,
  Class.\ Quant.\ Grav.\  {\bf 28}, 245020 (2011);
  R.~Ferraro, F.~Fiorini,
  Phys.\ Rev.\ D {\bf 84}, 083518 (2011);
  M.~H. Daouda, M.~E.~Rodrigues and M.~J.~S.~Houndjo,
  Eur.\ Phys.\ J.\ C {\bf 71}, 1817 (2011);
  M.~H.~Daouda, M.~E.~Rodrigues and M.~J.~S.~Houndjo,
  Eur.\ Phys.\ J.\ C {\bf 72}, 1890 (2012);
  P.~A.~Gonzalez, E.~N.~Saridakis and Y.~Vasquez,
  JHEP {\bf 1207}, 053 (2012);
  S.~Capozziello, P.~A.~Gonzalez, E.~N.~Saridakis and Y.~Vasquez,
  JHEP {\bf 1302}, 039 (2013);
  K.~Atazadeh and M.~Mousavi,
  Eur.\ Phys.\ J.\ C {\bf 72}, 2272 (2012).


\bibitem{Pereira}
R. Aldrovandi and J. G. Pereira, {\it Teleparallel Gravity: An Introduction},
Springer, Dordrecht (2013).



\bibitem{Maluf}
  J.~W.~Maluf,
  J.\ Math.\ Phys.\ \textbf{35} (1994) 335;
  H.~I.~Arcos and J.~G.~Pereira,
  Int.\ J.\ Mod.\ Phys.\ D \textbf{13}, 2193 (2004).


\bibitem{Maluf:2013gaa}
  J.~W.~Maluf,
  Annalen Phys.\  {\bf 525}, 339 (2013).


\bibitem{LandauLif} L.~D.~Landau and E.~M.~Lifschitz, {\it The Classical Theory of
Fields}, Addison-Wesley, Reading, MA (1971).





\bibitem{Davis:2007ta}
  S.~C.~Davis,
  arXiv:0709.4453 [hep-th];
  B.~Eynard and N.~Orantin,
  arXiv:0705.0958 [math-ph];
  A.~De Felice and S.~Tsujikawa,
  Phys.\ Rev.\ D {\bf 80}, 063516 (2009);
  A.~Jawad, S.~Chattopadhyay and A.~Pasqua,
  Eur.\ Phys.\ J.\ Plus {\bf 128}, 88 (2013).


\bibitem{muller-hoissen} F. Mueller-Hoissen and J. Nitsch,
Phys. Rev. D {\bf 28}, 718 (1983).


\bibitem{zanelli} A. Mardones and J. Zanelli,
Class. Quant. Grav. {\bf 8}, 1545 (1991);
O. Chandia and J. Zanelli, Phys. Rev. D {\bf 55}, 7580 (1997).


\bibitem{uscosmology}
  G.~Kofinas and E.~N.~Saridakis,
  arXiv:1408.0107 [gr-qc].



\end{thebibliography}
\end{document}